\documentclass[12pt,thmsa]{article}
\usepackage{amssymb}
\usepackage{epsf}
\usepackage{sw20nhj}

\input{tcilatex}

\normalsize

\begin{document}

\author{R. B. Fiorito \\
Catholic University of America, Washington, DC 20064\\
email : rfiorito@rocketmail.com \and D. W. Rule \\
Carderock Division, Naval Surface Warfare Center, W. Bethesda, MD 20817}
\title{Diffraction Radiation Diagnostics for Moderate to High Energy Charged
Particle Beams}
\maketitle

\begin{abstract}
\begin{quotation}
Diffraction radiation (DR) is produced when a charged particle passes
through an aperture or near a discontinuity in the media in which it is
traveling. DR is closely related to transition radiation (TR), which is
produced when a charged particle traverses the boundary between media with
different dielectric constants. In contrast to TR, which is now extensively
used for beam diagnostic purposes, the potential of DR as a
non-interceptive, multi-parameter beam diagnostic remains largely
undeveloped. For diagnostic measurements it is useful to observe backward
reflected DR from an circular aperture or slit inclined with respect to the
beam velocity. However, up to now, well founded equations for the
spectral-angular intensities of backward DR from such apertures have not
been available. We present a new derivation of the spectral angular
intensity of backward DR produced from an inclined slit for two orientations
of the slit axis, i.e. perpendicular and parallel to the plane of incidence.
Our mathematical approach is generally applicable to any geometry and
simpler than the Wiener Hopf method previously used to calculate DR from
single edges. Our results for the slit are applied to the measurement of
orthogonal beam size and divergence components. We discuss the problem of
separating the simultaneous effects of these beam parameters on the angular
distribution of DR and provide solutions to this difficulty. These include
use of the horizontal and vertical polarization components of the radiation
from a single slit and interferences from two inclined slits. Examples of DR
diagnostics for a 500 MeV beam are presented and the current limitations of
the technique are discussed.
\end{quotation}
\end{abstract}

\section{Introduction}

The new generation very high power radiation devices such as short
wavelength free electron lasers and single pass spontaneous emission (SASE)
devices for the generation of intense uv and x-ray beams requires the
development of linear accelerators which produce optical quality charged
particle beams. The high power, small size and low emittance of these beams
present an enormous challenge for both the diagnostic measurement of beam
parameters and the accurate positioning and control of these beams.
Conventional screens or other interceptive probes are incompatible with the
operation of such accelerators in many instances. Non interceptive devices
such as wall monitor arrays, give limited beam information. Other types of
monitors, such as synchrotron monitors, while non-invasive, cannot be used
in linear beam line geometries. Hence the development of low cost, compact,
nondestructive monitors capable of measuring multiple beam parameters would
be very useful for many high power beam applications.

We have investigated how the properties of diffraction radiation (DR) can be
used to measure beam divergence, energy, position, transverse beam size and
emittance. DR devices such as slits or circular apertures through which the
beam passes offer minimal perturbation to the beam, respond rapidly to
changes in beam parameters and are inherently compact, so that they can be
implemented at many places in the beam line.

Diffraction radiation (DR) is produced when a charged particle passes
through an aperture or near an edge or interface between media with
different dielectric properties. DR is closely related to transition
radiation (TR), which is produced when a charged particle traverses the
boundary between media with different dielectric constants. DR has been
studied theoretically since the early 1960's, [1--5]. However, in contrast
to TR, which has been well investigated experimentally and is now used
extensively for beam diagnosis, we are aware of only two experimental
studies of DR in the literature [6],[7]. In only one of these investigations
[6]\ was DR actually used as a beam parameter diagnostic, i.e. to measure
the beam bunch length. Furthermore, in this study forward and 180$^{0}$
backward directed DR and TR were observed simultaneously. This situation
complicated the comparison of measured to theoretically predicted DR
properties. Therefore, to avoid such difficulties and for experimental
convenience, it is advantageous to observe the radiation from an inclined
aperture, e.g. a slit inclined at 45$^{0}$, which produces DR both in the
forward direction and in the direction of specular reflection of the virtual
photons associated with the particle, i.e. at 90 degrees with respect to the
beam velocity.

The detailed nature of the far field angular distribution (AD) of \textit{%
forward} directed DR has been derived and discussed by Ter-Mikaelian [4]\
for an electron normally incident on a circular aperture and a slit in an
infinite, thin plane. The fields and AD of DR from an electron incident on a
single edge (half plane) inclined at an arbitrary angle with respect to the
velocity vector of the particle have also been derived and discussed in
Refs. [2], [4] and [8]. Although some authors have surmised that reflected
DR from an inclined circular aperture or slit has properties similar to
reflected TR [9-10], a derivation, from first principles, of the
spectral-angular distribution of reflected DR from such apertures has not
been previously available in the literature.\ We present here solutions for
the fields and the horizontally and vertically polarized intensities of DR
from an inclined slit for two orientations of the slit edge, i.e.parallel
and perpendicular with respect to the plane of incidence.

Previous theoretical investigations of forward directed DR generated from
electron beams passing through circular apertures and slits [9-12] have
examined the potential use of the far field angular distribution (AD) to
diagnose beam properties such as position with respect to the center of the
aperture, beam size, trajectory angle, divergence and energy. \ The
potential advantage and difficulty in using the AD of DR for beam
diagnostics is that it depends on the beam position (size) as well as the
beam divergence and energy. In contrast, the AD of TR depends only on the
beam divergence and energy. The challenge for the application of the AD of
DR as a diagnostic then is to devise a means to separate out the competing
effects of divergence and beam size on this observable.

We have developed a number of ways to address and solve this problem. We
demonstrate here how the measurements of the horizontal and vertical
polarization components of backward reflected DR from a single slit can be
used to separate out the effects of beam size perpendicular to the slit edge
and the component of divergence parallel to the slit edge. Furthermore, we
show how the measured AD's for two different orthogonal orientations of\ the
slit axis, can be used to determine the components of beam size and
divergence in two orthogonal directions, i.e. the directions perpendicular
and parallel to the plane of incidence. In addition we suggest how the
interferences from two slits can be used to make a sensitive measurement of
divergence alone.

\section{Diffraction Radiation Properties}

When a charged particle or ensemble of particles passes through an aperture
or near an edge at a distance $d$, appreciable DR is generated when the
condition: $d\lesssim \gamma \bar{\lambda}$ is fulfilled, where $\bar{\lambda%
}=\lambda /2\pi $, and $\lambda $ is the observed wavelength. Table I.
presents typical wavelengths of DR observed from a slit whose width $a=2d$ $%
=2mm$ for a variety of lepton and hadron beams. Table I. shows, for example,
that DR is produced in the optical part of the spectrum (i.e. IR to visible)
for electrons with energies in the range of 0.5 - 5 GeV.\medskip

\begin{center}
\textbf{Table I.}

\textbf{Typical wavelengths of diffraction radiation generated from a 2-mm
wide slit for various accelerators\medskip }

\begin{tabular}{|c|c|c|c|}
\hline
E(GeV) & $\gamma $ & $\lambda (\mu m)$ & Accelerator \\ \hline
0.5 & 10$^{3}$ & 6.30 & APS \\ \hline
5.0 & 10$^{4}$ & 0.630 & CEBA \\ \hline
50.0 & 10$^{5}$ & 0.063 & SLAC \\ \hline
250.0 & 250 & 25.00 & RHIC \\ \hline
10$^{3}$ & 10$^{3}$ & 6.30 & FNAL \\ \hline
\end{tabular}
\medskip
\end{center}

When the observed DR wavelength is much smaller than the transverse or
longitudinal size of a charged particle bunch, incoherent DR is produced
independently from each particle and the total intensity $I\sim N$, the
number of particles in the bunch. For wavelengths comparable to the bunch
dimensions coherent DR is observed with an intensity $I\sim N^{2}$. \ The DR
diagnostic methods we will present can utilize either incoherent or coherent
DR. However, for simplicity and to illustrate and emphasize the basic
methodology we will concern ourselves here only with the analysis of
incoherent DR. Furthermore, since we have discussed DR diagnostics using
circular apertures elsewhere [10], we will focus our attention in this paper
to the generation of DR from slits. An extended analysis of coherent DR from
a slit will be presented in a future work.

\section{Theory of Backward Diffraction Radiation from an Inclined Slit}

\subsection{\protect\normalsize \ Parallel orientation of slit with respect
to the plane of incidence}

A relativistic charged particle passing through a slit in a conducting
screen emits diffraction radiation in the forward and backward directions,
the latter in the direction of specular reflection of the virtual photons
associated with the relativistic particle. The peak intensity of diffraction
radiation, like transition radiation, occurs at an angle $\theta \sim \gamma
^{-1}$, where $\theta $ is measured from the direction of the particle's
velocity vector $\mathbf{v}$\textbf{\ }for forward DR, and from the
direction of specular reflection for backward DR.

Figure 1A. shows a side view, and Figure 1B. a top view, of the cone of \
backward diffraction radiation emitted by a charge passing through a slit
which is inclined at the angle $\Psi $ with respect to the velocity vector
and oriented such that its edge is horizontal or parallel to the plane of
incidence (\textbf{n, v} plane), where \textbf{n} is the unit vector normal
to the plane of the slit, i.e. the $(x^{\prime },y^{\prime })$ plane.
Additionally, the particle passes through the slit with an offset $\delta $
in the vertical $(y^{\prime })$ direction measured from the center of the
slit which is defined by the $x^{\prime }$ axis. For simplicity, we have not
drawn the forward directed cone of DR which is also generated in the
direction \textbf{v}. \ Figures 1A. and 1B. show the coordinate system $%
(x,y,z)$ used to describe the radiation fields associated with the wave
vector \textbf{k} observed in the backward direction (i.e. directed along
the $z$ axis). The slit is assumed to be an infinitely thin conducting
screen with infinite extent in the $x^{\prime }$ and $y^{\prime }$
directions beyond the slit aperture.

Our calculation of the backward reflected DR fields employs three different
concepts (a) the Huygens-Fresnel diffraction approach, following the
treatment given by Ter-Mikaelian [4]\ for the case of forward DR, (b) the
concept of backward reflected pseudo photons, used by Wartski [13] to
calculate reflected transition radiation and (c) a form of Babinet's
principle. Our approach is quite different and mathematically much less
complex than the exact Wiener Hopf method used by previous researchers to
calculate backward DR from a single edge [2,4]. In addition, our method is
generally applicable to other geometries where the Wiener Hopf technique is
very difficult to implement.

Consider a thin reflecting metal screen $S_{\infty }$with infinite extent. A
charged particle crossing this surface produces the well known forward and
backward transition radiation. \ Compare this to the case of Fig. 1., where
the slit has an area $S_{1}$ and the conducting screen has an area $S_{2}$.
\ Babinet's principle states that the transition radiation (TR) field from
the infinite conducting screen, $S_{\infty }$, can be obtained from a
Huygen's Fresnel integral over the infinite area $S_{\infty }=S_{1}+S_{2}$,
so that the TR field can be expressed as $\overrightarrow{E}_{\infty }=%
\overrightarrow{E}_{1}+\overrightarrow{E}_{2}$ where $\overrightarrow{E}_{1}$
and $\ \overrightarrow{E}_{2}$ are those parts of the field which are
obtained from a Huygen's Fresnel integral over the areas $S_{1}$ and $S_{2}$%
, respectively [8]. \ If material of area $S_{1}$ is removed, then the field 
$\overrightarrow{E}_{2}=\overrightarrow{E}_{\infty }-\overrightarrow{E}_{1}$%
, where $\overrightarrow{E}_{2}$ is the desired backward DR field. Note that
if the area $S_{1}\rightarrow 0$, \ $\overrightarrow{E}_{2}\rightarrow 
\overrightarrow{E}_{\infty }$ , and the DR field approaches that of TR.

\vspace{1pt}The incident field in the Huygens-Fresnel integral is the field
associated with the relativistic particle,$\overrightarrow{\text{ }E_{i}}$ .
Using Ter-Mikaelian's approach to calculate the diffraction radiation field
and Wartski's method of calculating backward reflected radiation, we can
express the $x$ and $y$ components of the radiation field as 
\begin{equation}
E_{x,y}(k_{x},k_{y})=r_{\parallel ,\perp }(\omega ,\Psi )\frac{1}{4\pi ^{2}}%
\iint E_{ix,iy}(x,y)e^{-i\overrightarrow{k}\cdot \overrightarrow{\rho }}dxdy
\label{1}
\end{equation}
where $E_{ix,iy}$ are the components of the field of the incoming charged
particle and $k_{x}$ and $k_{y}$ are the components of the wave vector $%
\mathbf{k}$ in the plane normal to the direction $\mathbf{z}$ shown in
Figure 1A. and 1B.\ As is the case for TR, there are generally three
components to the radiation: direct, reflected and transmitted. However, for
an opaque, highly conducting screen, the dominant term for backward TR or DR
production is the reflected component. The fields $E_{x}$ and $E_{y}$ for
backward TR and DR are thus proportional to the Fresnel reflection
coefficients $r_{\parallel }$ and $r_{\perp }$, respectively. In Eq. (1),
the integral over the area $S_{2}$ is done in terms of the coordinates $(x,$ 
$y)$ of Figures 1 and 2. The far field reflected spectral intensity of DR is
found by integrating the flux of pseudo photons reflected off the surface $%
S_{2}$, which pass through the plane that is perpendicular to the direction $%
\mathbf{z}$, i.e. the direction of specular reflection of incident pseudo
photons having a wave vector\textbf{\ }$\mathbf{k,}$ which is parallel or
closely parallel to the velocity vector $\mathbf{v}$. The phase term in Eq.
(1), which is a function of $\overrightarrow{\rho }=(x^{\prime },y^{\prime
}) $, a vector lying in the plane of the screen, can be related to the
coordinates $(x,y)$ by the transformations $x=x^{^{\prime }}\sin (\Psi )$, $%
y=y^{^{\prime }}$, and the differential elements of area by $%
dxdy=dx^{^{\prime }}dy^{^{\prime }}\sin (\Psi )$. The detailed solution of
the integral in Eq. (1) is presented in the Appendix.

The results are:

\begin{equation}
E_{x}(k_{x},k_{y})=\frac{ier_{\parallel }(\omega ,\Psi )}{4\pi ^{2}v}\frac{%
\overline{k}_{x}}{f}\left[ \frac{1}{(f-ik_{y})}e^{-a_{1}(f-ik_{y})}+\frac{1}{%
(f+ik_{y})}e^{-a_{2}(f+ik_{y})}\right] ,  \label{2}
\end{equation}
which is identical to Eq. (31.18) of Ref. [4], except for the reflection
coefficient $r_{\parallel }(\omega ,\Psi )$, and the scalar quantity $%
\overline{k}_{x}\equiv k_{x}-[k/(2\gamma ^{2})]\cot \Psi $, \ and

\begin{equation}
E_{y}(k_{x},k_{y})=\frac{er_{\perp }(\omega ,\Psi )}{4\pi ^{2}v}\left[ \frac{%
1}{(f-ik_{y})}e^{-a_{1}(f-ik_{y})}-\frac{1}{(f+ik_{y})}e^{-a_{2}(f+ik_{y})}%
\right] ,  \label{3}
\end{equation}
which is identical to Eq. (31.19) of Ref. [4], except for the factor $%
r_{\perp }(\omega ,\Psi )$. Here $a_{1,2}=a/2\pm \delta $, $f=[\overline{k}%
_{x}^{2}+\omega ^{2}/(v^{2}\gamma ^{2})]^{1/2}$, $v=\left| \mathbf{v}\right| 
$, $k_{x}=k\sin \theta _{x}$, $k_{y}=k\sin \theta _{y}$ and $\theta _{x,y}$
are the projected angles of the vector\textbf{\ }$\mathbf{k}$ into the $x,z$
and $y,z$ planes, respectively.

\ The horizontally polarized intensity (i.e. parallel to the slit edge) and
vertically polarized intensity (i.e. perpendicular to the slit edge), which
are observed in the plane perpendicular to the direction of specular
reflection, are defined in terms of the $x$ and $y$ components of the
fields: 
\begin{eqnarray}
\frac{d^{2}N_{horiz.}}{d\omega d\Omega } &=&\frac{2\pi ^{2}k^{2}c}{\hbar
\omega }\left| E_{x}\right| ^{2}  \label{4} \\
&=&\left| r_{\parallel }\right| ^{2}\frac{\alpha k^{2}\overline{k}%
_{x}^{2}e^{-af}}{4\pi ^{2}\omega f^{2}(f^{2}+k_{y}^{2})}[\cosh (2f\delta
)+\sin (ak_{y}+\Phi (k_{x},k_{y}))]  \nonumber
\end{eqnarray}
and

\begin{eqnarray}
\frac{d^{2}N_{vert.}}{d\omega d\Omega } &=&\frac{2\pi ^{2}k^{2}c}{\hbar
\omega }\left| E_{y}\right| ^{2}  \label{5} \\
&=&\left| r_{\perp }\right| ^{2}\frac{\alpha k^{2}}{4\pi ^{2}\omega }\frac{%
e^{-af}}{(f^{2}+k_{y}^{2})}[\cosh (2f\delta )-\sin (ak_{y}+\Phi
(k_{x},k_{y}))]  \nonumber
\end{eqnarray}
respectively, where

\begin{equation}
\Phi (k_{x},k_{y})=\sin ^{-1}[(f^{2}-k_{y}^{2})/(f^{2}+k_{y}^{2})]=\cos
^{-1}[-2fk_{y}/(f^{2}+k_{y}^{2})]\text{,}  \label{6}
\end{equation}
and $\alpha \simeq \frac{1}{137}$ is the fine structure constant.

The equations for vertically polarized \textit{forward} diffraction
radiation intensity, first obtained by Ter-Mikaelian [4] and later rewritten
and presented in Ref. [9], are both equivalent to Eq. (5 ) without the
factor $\left| r_{\perp }\right| ^{2}$. Note, however, that the equations
for phase term, $\Phi (k_{x},k_{y})$ presented in Ref. [9]\ are incorrect.
The correct relationships are given in Eq. (6).

\subsubsection{Limiting Forms {\protect\normalsize for Backward DR at 45}$%
^{0}$}

Under the usual small angle approximation for{\normalsize \ }observation of
TR and DR from relativistic particles, i.e. $v\approx c,\theta \sim \gamma
^{-1}\ll 1$, the quantities $\overline{k_{x}}\approx k_{x}\approx k\theta
_{x}$, $k_{y}\approx k\theta _{y}$, and $k_{z}\approx k$. We then can
consider several limiting forms for backward DR.

(1) \textit{Single Edge Limit}\textbf{:} When either $a_{1}$ or $a_{2\text{ }%
}\rightarrow \infty $ , the sinusoidal term in Eqs. (4) and (5), which
represents the interference between the intensities from each edge of the
slit, disappears and expressions for the horizontal and vertical intensities
of DR from a single edge are obtained. The addition of the horizontal and
vertical intensities taken in this limit produces an expression for the
total DR intensity which is identical to that obtained using the exact
Wiener Hopf approach (see Ref. [8], Eq. (14)). The correspondence of the
results of our calculations to those calculated using exact theory is a
strong confirmation of the validity of our method for calculating backward
reflected DR.

(2) \textit{Small Displacement Limit}: When $\delta ,$ the particle
displacement from the center of the slit, is small, i.e. $\delta \ll $ $%
\gamma \bar{\lambda}\sim a$, the hyperbolic cosine in Eqs. (4) and (5) can
be expanded in a Taylor series. Retaining terms only up to second order, and
using the small angle approximations given above, Eqs. (4) and (5) become: 
\begin{eqnarray}
\frac{d^{2}N_{horiz.}}{d\omega d\Omega } &=&\left| r_{\parallel }\right| ^{2}%
\frac{\alpha }{4\pi ^{2}\omega }\gamma ^{2}\frac{X^{2}}{(1+X^{2})}\frac{%
e^{-R(1+X^{2})^{1/2}}}{(1+X^{2}+Y^{2})}  \label{7} \\
&&{\small \cdot }[1+2(\frac{\delta }{\gamma \bar{\lambda}}%
)^{2}(1+X^{2})+\sin (RY+\Phi ^{\prime }(X,Y))]  \nonumber
\end{eqnarray}
and

\begin{eqnarray}
\frac{d^{2}N_{vert.}}{d\omega d\Omega } &=&\left| r_{\perp }\right| ^{2}%
\frac{\alpha }{4\pi ^{2}\omega }\gamma ^{2}\frac{e^{-R(1+X^{2})^{1/2}}}{%
(1+X^{2}+Y^{2})}  \label{8} \\
&&{\small \cdot [}1+2(\frac{\delta }{\gamma \bar{\lambda}}%
)^{2}(1+X^{2})-\sin (RY+\Phi ^{\prime }(X,Y))]  \nonumber
\end{eqnarray}
where we have introduced the new variables $X=\gamma \theta _{x}$ and $%
Y=\gamma \theta _{y}$, the $x$ and $y$ projected angles scaled in units of $%
\gamma ^{-1}$, $R\equiv a/\gamma \bar{\lambda}$ and the reduced phase term 
\begin{equation}
\Phi ^{^{\prime }}(X,Y))=\sin ^{-1}[(1+X^{2}-Y^{2})/(1+X^{2}+Y^{2})]=\cos
^{-1}[-2(1+X^{2})^{1/2}Y/(1+X^{2}+Y^{2})].  \label{9}
\end{equation}

(3) \textit{TR Limit}: Additionally, when $R<<1$ and $\delta <<\gamma \bar{%
\lambda}$ , Eqs. (7) and (8) each reduce to one half the intensity of TR,
which is the expected correct limit. In this regime the particle radiates as
if the slit in the screen were absent.

\subsection{\protect\normalsize Perpendicular orientation of the slit with
respect to the plane of incidence}

Figures 2A. and 2B. illustrate the case in which the slit is oriented such
that the edge is perpendicular to the plane of incidence, i.e. the plane
containing \textbf{v }and \textbf{n} the normal to the plane of the screen.
In this case, the offset $\varepsilon $ of the particle velocity vector from
the center of the slit is in the direction $Y^{\prime }$ and the radiation
fields $E_{x}$ and $E_{y}$ can be shown to be the same as those given above
in Eqs. (2) and (3) but with $a_{1}\rightarrow a_{1}\sin \Psi $, $%
a_{2}\rightarrow a_{2}\sin \Psi $ and $\delta \rightarrow \varepsilon \sin
\Psi $. Also,\ $E_{x}\varpropto r_{\perp }(\omega ,\Psi )$ and $%
E_{y}\varpropto r_{\parallel }(\omega ,\Psi )$ since, for this orientation
of the slit, $\overrightarrow{E}_{x}$ is perpendicular and $\overrightarrow{%
E_{y}}$ is parallel, respectively, to the plane of incidence. Then $E_{x}$
and $E_{y\text{ }}$take the same forms as Eqs. (2) and (3), with the
coefficients $r_{\perp }$ and $r_{\parallel }$ interchanged. Similarly the
vertical and horizontal intensity components take on the same forms as Eqs.
(4) and (5), with the coefficients $\left| r_{\perp }\right| ^{2}$ and $%
\left| r_{\parallel }\right| ^{2}$ interchanged.

In Ref. [8] it has been shown that an exact calculation of reflected DR
produced by a particle traveling with a velocity with a parallel component
along the direction of the edge of a semi-infinite screen [2] reduces, in
the limit $\gamma >>1$, to the same result as that of a particle whose
velocity vector is purely perpendicular to the edge (see Eqs.(7) and (14) of
Ref.[8]). The former situation corresponds to the geometry of Figure 1. as
discussed above in Section 3.1; the latter situation corresponds to the
geometry of Figure 2., which is discussed in this Section. As in the
parallel case, the single edge results for the geometry of Fig. 2. are
completely reproduced using our approach in the limit $a_{1}$ or $a_{2\text{ 
}}\rightarrow \infty $. Furthermore, in the perpendicular incidence case the
small angle, relativistic limit produces the same forms for the horizontal
and vertical intensities as the parallel incidence case, i.e. Eqs. (7) and
(8), but with the coefficients $\left| r_{\perp }\right| ^{2}$ and $\left|
r_{\parallel }\right| ^{2}$ interchanged. In addition, when $R<<1$, the
horizontal and vertical intensities each go to the same TR limit as in the
parallel incidence case. These results provide further evidence of the
soundness of our theoretical approach.

For the sake of simplicity and to illustrate the main results of our
analysis of DR from slits, we will present numerical examples only for the
case of the slit oriented with its edge\textit{\ parallel} to the plane of
incidence . However, numerical results can easily be obtained for the
perpendicular orientation by simple substitution of variables as listed
above.

\section{Discussion and Computational Results}

\subsection{General properties of DR}

Eqs. (5) and (6) indicate that the highest yield of backward DR observed in
the IR to visible part of the spectrum will be generated when the surface of
the slit is a highly conducting mirrored surface. In the case $\left|
r_{\perp }\right| ^{2}\simeq $ $\left| r_{\parallel }\right| ^{2}\simeq 1$.
and for $\gamma >>1$, Eqs. (7) and (8) show that the polarization components
are the same as forward DR. These equations also reveal some interesting
characteristics of DR from a slit, which have not been fully discussed in
previous studies, and which can be easily visualized with the help of three
dimensional plots of intensity observed in the $X,Y$ plane.

Figs. 3A. and 3B. show the horizontal and vertical components, respectively,
of the intensity of DR for a value of the ratio $R=a/(\gamma \bar{\lambda}%
)=0.5$. Similarly, Figs. 4A. and 4B. show the component intensities for $%
R=2.0$ . The intensity axis $(Z)$ in these Figures is scaled in proportion
to the corresponding intensity component of transition radiation. For
example, Fig. 3A indicates that the peak intensities of horizontal and
vertical DR, respectively, are about 25\% and 80\% that of TR , when $R=0.5.$
Figures 3. and 4. also show that diffraction fringes are present in both
polarization components, but are only observed in the $Y$ direction
(perpendicular to the slit edge) for both the orientations shown in Figs. 1
and 2. This can be deduced directly by an examination of the term $\sin
(RY+\Phi ^{\prime }(X,Y))$ in Eqs. (7) and (8), i.e. the frequency of
oscillation is determined solely by the value of $k_{y}$ component of the
wave vector represented through the variable $R.$ Secondly, the inteferences
in the horizontal and vertical intensities are $180^{0}$ out of phase. This
is evident by the presence of the difference in the sign of the sinusoidal
terms in each component. Thus, in general, the maxima of the vertical
component along the $Y$ axis occur at the minima of the horizontal component.

One further observes that the equation for horizontal intensity, Eq.(7)
contains the factor $X^{2}/(1+X^{2})$ , which is not present in Eq.(8) for
the vertical intensity. This term, which is similar to the form of TR,
forces the horizontal component to have a null at $X=0$ for every value of $Y
$. A comparison of the ratio of vertical to horizontal intensities reveals
that the vertical component is larger than the horizontal component, and
that ratio becomes smaller as $R$ decreases approaching unity as $%
R\rightarrow 0$ , i.e. the transition radiation limit. Also, as $R$ becomes
large the DR intensity drops off exponentially. Thus the value of $R$ should
be kept to a minimum to maximize the observed intensity of DR. We have
chosen the values $R=0.5$ and $2.0$ to show that measurable horizontal and
vertical intensity levels can be achieved for reasonable experimental
conditions and to demonstrate that the ratio of vertical to horizontal
intensity depends on the value of $R$. For example, a value of $R=2$ applies
to the case when the beam energy $E_{b}=500$ $MeV$, the slit width $a=2mm$
and the wavelength $\lambda $ $\simeq 3\mu m$. By change of the parameters $a
$ and $\lambda $, or by use of different values of $R$ , other workable
combinations are possible for a given beam energy. Note that as the beam
energy increases, a smaller value of $R$ can be achieved for the same
wavelength and slit size. Since the intensity of DR, like TR, is
proportional to $\gamma ^{2}$ (see Eqs. (7) and (8)), DR becomes
progressively more intense as hence easier to detect as the beam energy
increases.

In general, the sum of horizontal and vertical polarization intensity
components will be observed by a detector in the $X,Y$ plane. Thus, if the
angular distribution of DR is imaged, e.g. by using a camera with a lens
focussed at infinity, the sum of horizontal and vertical distributions will
be superimposed. However, either distribution can be imaged separately by
placing a rotatable polarizer in front of the lens. We will show that each
of these intensity components provides information about the beam size and
divergence in each of two orthogonal directions and, furthermore, that all
of these beam parameters can be measured by proper analysis of the
horizontally and vertically polarized intensities.

\subsection{\protect\normalsize Effect of Beam Parameters on the Angular
Distribution of DR}

\subsubsection{Beam Size}

The above discussion shows that both the horizontal and vertical DR
spectral-angular distributions are functions of the beam position $\delta $
relative to the center of the slit. This effect has been previously
considered [9,11] as a possible diagnostic of the beam position and size. In
Ref. [9] it was shown that the effects of beam size and offset on the
angular distribution of DR are the same. Also, in Ref. [11] it was shown
that the DR intensity is a minimum for a beam centered in the slit, i.e.
zero offset. Since the beam can be centered by monitoring and steering the
beam to minimize the total intensity, the offset can be nullified or
otherwise determined, for example, by use of a standard wall current
monitor. The remaining effects on the angular distribution of the DR will
then be the beam size in the direction perpendicular to the slit edge and
the two orthogonal components of the beam divergence.

To determine the effect of beam size on the intensity components, we assume
that the beam centroid has been centered in the slit by means of one of the
previous methods described above and that any particle within the finite
beam spatial distribution has the same effect as the offset of a single
particle. We further assume that the beam has a separable spatial
distribution given by $S=$ $S_{1}(\delta )\cdot S_{2}(\varepsilon )$, where $%
\delta $ and $\varepsilon $ are in the directions $y^{\prime }$ and $%
x^{\prime }$, respectively. For the parallel orientation of the slit, $%
\delta $ and $\varepsilon $ are perpendicular and parallel, respectively, to
the slit edge. Integration of the distribution $S$ over $\delta $ and $%
\varepsilon $ produces the average beam sizes $\left\langle \delta
\right\rangle $ and $\left\langle \varepsilon \right\rangle $. Because of
the simple dependence of the above expressions for horizontal and vertical
intensity on the variable $\delta $, an integration of $S$ over the
intensity components is mathematically equivalent to a simple replacement
the variable $\delta $ with its average value $\left\langle \delta
\right\rangle $. To simplify the notation in the remainder of the text we
will identify and refer to $\delta $ and $\varepsilon $ as the othogonal rms
beam size components, which are perpendicular and parallel, respectively, to
the plane of incidence.

\subsubsection{ Beam Divergence}

Ref. [9] has examined the effect of beam size perpendicular to the edge of a
slit on the vertical intensity component of forward DR in the absence of
divergence. A discussion of the effect of beam size on the horizontal
component was dismissed because of its much lower intensity in comparison to
the vertical intensity - an effect which was due to the rather large value
of $R=2\pi $ used in the analysis. However, for any realistic beam, the
divergence is not negligible and affects both the horizontal and vertical
intensities of DR. Furthermore, in general, the horizontal component of DR
is not negligible in comparison to the vertical component and, as we will
show, this component provides important information on the component of beam
divergence parallel to the slit edge. It is therefore important to analyze
both the effect of beam size and divergence on the vertical and horizontal
intensity components to fully assess the diagnostic potential of DR.

The effect of divergence must be taken into account by performing a two
dimensional convolution of a distribution of particle trajectory angles
projected in the $X,Z$ and $Y,Z$ planes (e.g. separable Gaussian
distributions in $\theta _{x}$ and $\theta _{y}$). However, we have shown by
numerical calculation that the effect of the rms divergence $\delta ^{\prime
}$ on the \textit{horizontal} component of the DR intensity, and the effect
of the rms divergence $\varepsilon ^{\prime }$ on the \textit{vertical}
component is insignificant for $\delta ^{\prime },\varepsilon ^{\prime }$ $%
\lesssim $ 0.2, a value which is large by the standards of most high quality
accelerators. Then, to a good approximation, a one dimensional convolution
of a line scan of the horizontal or vertical intensity calculated either in
the plane defined by $Y=const.$ or $X=const.$ can in principal be used to
predict the effects of the beam divergences $\varepsilon ^{\prime }$ and $%
\delta ^{\prime }$, respectively. Inversely, these divergences can be
determined by fitting the convolved intensities to measured data.

\paragraph{Horizontal Intensity Component}

The effect of the  divergence $\varepsilon ^{\prime }$, the component
parallel to the incidence plane, on the horizontal intensity component can
be estimated by neglecting the beam size term. Consider the term $X^{2}$ in
the numerator of the horizontal component Eq.(7). Because of the presence of
this term, a convolution of the horizontal intensity component with a
distribution of particle angles $G(X,\varepsilon ^{\prime })$ will have the
maximum effect on this component. Furthermore, the effect of $\varepsilon
^{\prime }$ will be maximized near the center of the pattern where the
horizontal intensity goes to zero in the absence of divergence. One can
estimate this effect by setting the variable $Y$ in Eq. (7) equal to a
constant value $Y=\delta ^{\prime }=const.$, and noting the effect on a line
scan of the one dimensional convolution of the horizontal intensity with
e.g. the Gaussian distribution: $G(X,\varepsilon ^{\prime })=\frac{1}{\sqrt{%
2\pi \varepsilon ^{\prime 2}}}\exp (\frac{-X^{2}}{2\varepsilon ^{\prime 2}})$%
.

Figure 5. shows two such normalized scans of a 1D convolution of horizontal
intensity (Eq. (7)) with $G$ for two different values of the variable, $Y=$ $%
\delta ^{\prime }=0.2$ and $Y=0$. In this calculation, $R=2.0$ and $%
\varepsilon ^{\prime }=0.1$. As Fig. 5. shows, there is no discernible
difference between the two patterns. Then only a small error will be made by
setting the variable $Y$ equal to a constant and performing a one
dimensional convolution over $X$ to infer the effect of the divergence $%
\varepsilon ^{\prime }.$ Inversely, it is possible to measure the
divergence, $\varepsilon ^{\prime }$ from line scans of horizontally
polarized DR for a number of $Y$ values by comparing the measured intensity
scans to theoretically convolved line scans in the variable $X$ , which are
parameterized by the variable $\varepsilon ^{\prime }$.

Now consider the effect of the beam size $\delta $ measured perpendicular
the plane of incidence, which is also perpendicular to the slit edge for the
orientation shown in Fig. 1., on the horizontal intensity. Note that the
effect of beam size $\varepsilon $ or any displacement parallel to the slit
edge has no effect on this component. To compare the effect of the beam size 
$\delta $ with effect of the divergence $\varepsilon ^{\prime }$, consider a
line scan of the horizontal intensity component observed over a finite
bandwidth $(\Delta \omega )$ taken in the plane, $Y=0$. Eq. (7) then reduces
to: 
\begin{equation}
\frac{dN_{Horiz.}}{d\Omega }=\left| r_{\parallel }\right| ^{2}\frac{\alpha
\gamma ^{2}}{2\pi ^{2}}\frac{\Delta \omega }{\omega }\frac{X^{2}}{%
(1+X^{2})^{2}}e^{-R(1+X^{2})^{1/2}}{\small \cdot }[1+(\frac{\delta }{\gamma 
\bar{\lambda}})^{2}(1+X^{2})].  \label{10}
\end{equation}

A convolution of this expression with the Gaussian distribution, $%
G(X,\varepsilon ^{\prime })$, evaluated at $X=0$, to first order in $%
\varepsilon ^{\prime 2}$, gives 
\begin{equation}
\frac{dN_{Horiz.}}{d\Omega }\otimes G(X,\varepsilon ^{\prime })\propto
\varepsilon ^{\prime 2}[1+(\frac{\delta }{\gamma \bar{\lambda}})^{2}].
\label{11}
\end{equation}
Thus, while it is still present, the effect of the beam size $\delta $ on
the horizontal component will be much smaller than that of the divergence $%
\varepsilon ^{\prime }$. Figure 6. shows how a change in the divergence $%
\varepsilon ^{\prime }$ affects the horizontal intensity for a fixed value
of the beam size $\delta =300\mu $. As predicted from Eq.(11), the effect of 
$\varepsilon ^{\prime }$ is maximum near the origin. Conversely, the shape
of pattern shape in the vicinity of the origin can be used to separate out
and measure the divergence $\varepsilon ^{\prime }$.

Figure 7. shows the effect of the beam size, $\delta $ on a line scan of the
horizontal intensity (Eq. (12)) taken in the $X,Z$ plane, i.e. $Y=0$, when $%
\varepsilon ^{\prime }=0$ and $R=2$. Figure 7. shows that a change in $%
\delta $ has no effect on the horizontal component in the region near $X=0$,
but does affect the peak value of the intensity observed at $\left| X\right|
\approx 1$ and the fall off of intensity for $\left| X\right| >1$, for $%
\delta >50\mu $. Numerical calculations indicate similar variation of the
intensities with beam size and divergence for $R=0.5$. Therefore, for
simplicity, we will only present below numerical results for a single value
of the parameter $R=2$.

\paragraph{Vertical Intensity Component}

A procedure similar to the one described above can be used to show that the
effect of the divergence $\varepsilon ^{\prime }$ has a minimal effect on
the vertical intensity distribution. Figure 8. shows two scans of the
vertical intensity observed in Figure 4B in the planes $X=\varepsilon
^{\prime }=0.2$ and $X=0$. As similarly shown in Fig. 5., the difference in
the two scans is very small. Thus, it should be possible to use a one
dimensional convolution to infer the effect of the divergence $\delta
^{\prime }$, on line scans of vertically polarized DR (measured in the $Y$
direction) for a number of $X$ values.

By comparing the measured intensity scans to a set of one dimensional
convolutions in $Y$ parameterized by the divergence component $\delta
^{\prime }$, one would hope to be able to measure this quantity. However,
for the vertical intensity component the effect of the beam size $\delta $
is comparable to that of the divergence $\delta ^{\prime }$. To see this
consider an observation of the vertical intensity component (Eq. (8)) in the 
$Y,Z$ plane i.e. $X=0$. For a finite bandwidth ($\Delta \omega )$
measurement, Eq. (8) produces the vertical angular distribution 
\begin{eqnarray}
\frac{dN_{Vert.}}{d\Omega } &=&\left| r_{\perp }\right| ^{2}\frac{\alpha }{%
4\pi ^{2}}\frac{\Delta \omega }{\omega }\gamma ^{2}\frac{e^{-R}}{(1+Y^{2})}
\label{12} \\
&&\cdot [1+2(\frac{\delta }{\gamma \bar{\lambda}})^{2}+\frac{2Y\sin
(RY)-(1-Y^{2})\cos (RY)}{(1+Y^{2})}].  \nonumber
\end{eqnarray}

To compute the effect of the divergence $\delta ^{\prime }$ one must take
the convolution of Eq. (12) with a distribution over the variable $Y$, e.g.
the Gaussian distribution. $G(Y,\delta ^{\prime })=\frac{1}{\sqrt{2\pi
\delta ^{\prime 2}}}\exp (\frac{-Y^{2}}{2\delta ^{\prime 2}})$. We have
performed this computation numerically to see how the effects of beam size
and divergence compare. Figure 9. shows a vertical intensity scan taken at $%
X=0$, for a fixed value of the beam size $\delta =$200$\mu $ and $R=2.0$ for
several values of divergence $\delta ^{\prime }$. Figure 10. shows the
effect of a change in the beam size, $\delta $ on similar scans for the
divergence value $\delta ^{\prime }=0$. A comparison of Figs. 9. and 10.
clearly indicates that the effect of divergences $\delta ^{\prime }$ $<0.2$
will compete significantly with the effect of beam sizes $\delta \lesssim
300\mu $ on the angular distribution of the vertical intensity component.
Thus, in general, neither effect can be neglected and the two effects are
not separable for this component.

\subsection{Strategies for separating beam size and divergence effects}

\subsubsection{Use of slit oriented perpendicular to the plane of incidence}

In order to help separate out the competing effects of beam size and
divergence, we have considered two alternative strategies. One method is to
rotate the slit (or insert another into the beam line) so that the slit edge
is perpendicular to the plane of incidence (see Fig. 2A.). In this
configuration the horizontal intensity component of the DR, convolved with
the Gaussian function, $G(X,\delta ^{\prime })=\frac{1}{\sqrt{2\pi \delta
^{\prime 2}}}\exp (\frac{-X^{2}}{2\delta ^{\prime 2}}),$ can be used to
measure the divergence $\delta ^{\prime }$, which is now the component of
divergence measured parallel to the slit edge. For this orientation the
horizontal component is highly sensitive to $\delta ^{\prime }$ and only
weakly depend on the beam size $\varepsilon $ , which is now the component
of the beam size perpendicular to the slit edge. Evaluating the convolved
horizontal intensity at $X=0$, we obtain 
\begin{equation}
\frac{dN_{Horiz.}}{d\Omega }\otimes G(X,\delta ^{\prime })\propto \delta
^{\prime 2}[1+(\frac{\varepsilon }{\gamma \bar{\lambda}})^{2}].  \label{13}
\end{equation}
Eq.(13) is the direct analog of Eq.(11) above which, as shown in the
previous section, can be used to determine $\varepsilon ^{\prime }$. For the
perpendicular orientation of the slit edge, however, the center of the
horizontal AD pattern is sensitive to, and can be used to measure, the
divergence component $\delta ^{\prime }$.

The vertical DR intensity component for this orientation of the slit is: 
\begin{eqnarray}
\frac{dN_{Vert.}}{d\Omega } &=&\left| r_{\parallel }\right| ^{2}\frac{\alpha 
}{4\pi ^{2}}\frac{\Delta \omega }{\omega }\gamma ^{2}\frac{e^{-R}}{(1+Y^{2})}
\label{14} \\
&&\cdot [1+2(\frac{\varepsilon }{\gamma \bar{\lambda}})^{2}+\frac{2Y\sin
(RY)-(1-Y^{2})\cos (RY)}{(1+Y^{2})}],  \nonumber
\end{eqnarray}
which is the direct analog of Eq. (12). A one dimensional convolution of Eq.
(14) with the Gaussian distribution $G(Y,\varepsilon ^{\prime })=\frac{1}{%
\sqrt{2\pi \varepsilon ^{\prime 2}}}\exp (\frac{-Y^{2}}{2\varepsilon
^{\prime 2}})$, produces a pattern that is sensitive to the divergence $%
\varepsilon ^{\prime }$ as well as the beam size $\varepsilon $. Thus for
the perpendicular slit orientation the divergence $\varepsilon ^{\prime }$
and beam size $\varepsilon $ have comparable effects and it is not possible
to distinguish between them using the vertical intensity component alone.

\subsubsection{DR Interferometry}

The second strategy we have devised to separate out and measure divergence
is to use the interference of DR produced from two slits inclined at $45^{0}$
with respect to the beam velocity in a configuration which is the direct DR
analogy to a Wartski OTR interferometer[13]. In such a system forward DR
from the first slit reflects from the second slit surface and interferes
with backward DR generated from the second slit. The interferences produced
by the two intensities will be superimposed on the single slit DR intensity.
Figure 11. shows one possible configuration for a DR interferometer composed
of two slits oriented with their edges parallel to the plane of incidence.
The equation for interference DR, in analogy to TR, is obtained by
multiplying Eqs. (7) and (8) by an additional interference term which is due
to the difference in phase between forward and backward DR. The expressions
for the horizontal and vertical intensity components are of the form: 
\begin{equation}
\frac{dN_{Horiz,Vert}^{(I)}}{d\Omega }=4\frac{dN_{Horiz,Vert}^{(S)}}{d\Omega 
}\sin ^{2}(\frac{L}{2L_{V}})  \label{15}
\end{equation}
where the superscript $I$ refers to the two slit (interferometer), and $S$
to the single slit angular intensity distributions, respectively, $%
L_{V}\equiv \lambda /\pi (\gamma ^{-2}+\theta ^{2})$ is the coherence length
in vacuum for TR and DR, which respresents the distance over which the
particle's field and the TR or DR photon differ in phase by one radian and $L
$ is the path length between the slits.

For a fixed wavelength and energy the single slit term varies slowly with
the angle $\theta $. Then a convolution of Eq.(15) with a distribution of
beam angles will chiefly effect the interference term $\sin ^{2}(\frac{L}{%
2L_{V}})$. The interference fringe visibility is then a function of the
angular divergence and is independent of beam size effects. Thus it should
be possible to use DR interferences to obtain a measurement of the beam
divergence alone. Also as is the case with interference OTR, measurements of
the two polarized components of interference DR can be used to measure
orthogonal components of the beam divergence[14]. The interference fringes,
which are superimposed on the single slit angular distribution, provide
increased sensitivity to angular differences in the particle trajectory
angles and therefore smaller values of divergence ($\delta ^{\prime }$,$%
\varepsilon ^{\prime }\lesssim 0.05$) can be measured with an interferometer
than with a single slit alone.

Figures 12. and 13. show the vertical and horizontal components of
intensity, respectively, from a DR interferometer produced by an electron
beam with energy $E=500$ MeV, $R=2$, $\lambda $=3.2$\mu $m, $\delta $=200$%
\mu $ and $\delta ^{\prime }$=0.05. The separation distance $L=6L_{V}=3$
meters for this beam energy. In any DR interferometer the angular
distribution of the forward DR from the first foil will be partially cut off
by the presence of the second slit in such a configuration. If the path
length between the slits is large, the total angular field will only be cut
off in the center of the pattern by a small amount. Thus only a small
fraction of the interference patterns ($1/6\gamma $ for the parameter range
specified above) will be cut out. This effect is not shown in Figs. 12. and
13.

In contrast to the very weak effect of a small value of divergence, i.e. $%
\delta ^{\prime }=0.05,$ on single slit DR (see Figs. 7. and 9.), the effect
of this divergence on the DR interference fringe visibility is clearly
visible. Since the fringes are well modulated out to $Y\sim 4$, the cutoff
of the field of view due to the second aperture will have a negligible
effect on a measurement of the divergence, which uses the interference
pattern. With the help of a second camera or imager, a simultaneous
measurement of backward DR from the first slit can be made and thereby
provide addition single slit data for extracting the beam size.

At the present time the use of the angular distribution of DR for
diagnostics is practically limited to beams with moderate to high energies
by the vacuum coherence length $L_{V}\sim \gamma ^{2}\lambda $. The DR
produced by upstream sources such as beam line discontinuities or other
apertures will destructively interfere with backward DR generated from a
diagnostic aperture when the distance between the upstream source and the
aperture is much less than $L_{V}$. Also, the successful application of
interference DR to measure beam divergences $\delta ^{\prime },\varepsilon
^{\prime }\lesssim 0.05$ requires that the inter-aperture distance $L\gtrsim
5L_{V}$. Therefore, the diagnostic techniques described in this paper are
limited to observation wavelengths and beam energies which do not give rise
to an impracticably long coherence length for the accelerator facility being
used. The means to overcome this limit must be developed before the methods
described here can be applied to beams with very high energy.

\section{Summary}

We have developed a new method to calculate backward reflected diffraction
radiation from any type of aperture. Using this method we have derived the
equations for backward horizontal and vertically polarized spectral-angular
intensities of DR from a slit inclined at an arbitrary angle with respect to
the particle velocity. We have obtained results for two orthogonal
orientations of the slit edges, i.e. parallel and perpendicular to the plane
of incidence. The results of our calculations reduce in the single edge
limit precisely to those previously derived using the exact Wiener Hopf
method. However, in contrast to the Wiener Hopf approach, which has been
successfully used to calculate DR from one type of aperture only, i.e. the
single edge, our approach is readily applicable to any type of aperture and
is much simpler to employ.

From an analysis of backward DR for the two orientations of the slit
mentioned above, we have developed a number of strategies to determine four
unknown beam parameters of interest: the two orthogonal beam divergences, $%
\delta ^{\prime }$, $\varepsilon ^{\prime }$, which are perpendicular and
parallel to the incidence plane, respectively, and the two corresponding
orthogonal components of the beam size: $\delta ,\varepsilon $. These
strategies require the measurement of the horizontally and vertically
polarized intensities from a single slit or a double slit DR interferometer.
We have demonstrated that one dimensional convolved scans of the horizontal
intensities in either the $X=$ const. or $Y=$ const. plane can be used to
separately measure $\delta ^{\prime }$ and $\varepsilon ^{\prime }$,
respectively. With the divergences known, the corresponding beam size
components $\delta $ and $\varepsilon $ can be inferred. Since each of these
scans can be produced at multiple angles in the plane of observation, a
large amount of data is available which can be used to reduce the error in
the measurement of all the beam parameters of interest. The stategies we
have developed are useful for moderate to high energy lepton or hadron beams.

\section{Appendix}

To proceed to evaluate the integral presented in Eq. (1) we express $%
E_{ix}(x,y)$ by its Fourier transform, so that we have 
\begin{equation}
E_{x}(k_{x},k_{y})=\frac{ie}{8\pi ^{4}}\frac{1}{v}r_{\parallel }(\Psi )\iint 
\frac{k_{x}^{\prime }}{(k_{x}^{\prime 2}+k_{y}^{\prime 2}+\bar{\alpha}^{2})}%
e^{i(\overrightarrow{k^{\prime }}-\overrightarrow{k})\cdot \overrightarrow{%
\rho }}dk_{x}^{\prime }dk_{y}^{\prime }dxdy\text{.}  \label{16}
\end{equation}
Here, $\bar{\alpha}\equiv 1/(\beta \gamma \bar{\lambda})$, and the phase
term appearing in Eq. (16) above is 
\begin{equation}
(\overrightarrow{k^{\prime }}-\overrightarrow{k})\cdot \overrightarrow{\rho }%
=(k_{x}^{\prime }-\bar{k}_{x})x^{\prime }\sin \Psi +(k_{y}^{\prime
}-k_{y})y^{\prime }  \label{17}
\end{equation}
where 
\begin{equation}
\bar{k}_{x}\equiv k_{x}+(k_{z}-\overrightarrow{k^{\prime }}\cdot \frac{%
\overrightarrow{v}}{v})\cot \Psi \text{.}  \label{18}
\end{equation}
Note that $\overrightarrow{k^{\prime }}\cdot \overrightarrow{v}$ $=\omega $
, where $\omega $ is the frequency of the Fourier component of the field.
This phase term takes into account the variation in phase with position $%
\overrightarrow{\rho }$ of the $\overrightarrow{k^{\prime }}$ Fourier
component of the radiation field. For relativistic electrons, typically, $%
\gamma >>1$ and $\overrightarrow{k}$ is nearly along the direction $Z$ in
Fig. 1., which is the direction into which a ray along $\overrightarrow{v}$
would be reflected if the screen were replaced by a mirror. The vector $%
\overrightarrow{k}=\overrightarrow{k_{x}}+\overrightarrow{k_{y}}+%
\overrightarrow{k_{z}}\approx k(\theta _{x}\hat{x}+\theta _{y}\hat{y}+\hat{z}%
)$ (see the $x,y,z$ coordinate system in Fig. 1.), where $\theta ^{2}=\theta
_{x}^{2}+\theta _{y}^{2}$ and $\theta _{x}$,$\theta _{y}<<1$ . For $\Psi
=45^{0}$ , we have $\bar{k}_{x}\approx k(\theta _{x}-\gamma ^{-2}/2)\approx
k\theta _{x}$ . Therefore, in general, the radiation pattern is shifted by
an angle $\gamma ^{-2}/2<<1$ .

The $x$ and $y$ integrations are straight forward. The variable $y^{^{\prime
}}$ ranges from $-\infty $ to $a_{1\text{ }}$and $a_{2}$ to +$\infty $,
where the electron trajectory along $\overrightarrow{v}$ is taken to be a
distance $a_{1}$ from the lower edge and a distance $a_{2}$ from the upper
edge of the slit so that $a=a_{1}+a_{2}$ in Fig. 1. The $x$ integration
yields $2\pi \delta (k_{x}^{\prime }-\bar{k}_{x})$ so that the $%
k_{x}^{\prime }$ integral can be done trivially. Using $\bar{k}_{x}\approx
k_{x}$, when $\gamma >>1$ , we obtain 
\begin{equation}
E_{x}(k_{x},k_{y})=\frac{ie}{2\pi ^{2}}\frac{1}{v}r_{\parallel }(\Psi
)\left\{ \frac{1}{2\pi }[I_{1}+I_{2}]+\frac{k_{x}}{(k_{x}^{2}+k_{y}^{2}+\bar{%
\alpha}^{2})}\right\}  \label{19}
\end{equation}
where 
\begin{equation}
I_{1}=(-i)\int_{-\infty }^{+\infty }dk_{y}^{\prime }\frac{k_{x}}{%
(k_{x}^{2}+k_{y}^{2}+\bar{\alpha}^{2})}\frac{1}{k_{y}^{\prime }-k_{y}}%
e^{-i(k_{y}^{\prime }-k_{y})a_{1}},  \label{20}
\end{equation}
and 
\begin{equation}
I_{2}=(+i)\int_{-\infty }^{+\infty }dk_{y}^{\prime }\frac{k_{x}}{%
(k_{x}^{2}+k_{y}^{2}+\bar{\alpha}^{2})}\frac{1}{k_{y}^{\prime }-k_{y}}%
e^{i(k_{y}^{\prime }-k_{y})a_{2}}.  \label{21}
\end{equation}

Now we can write 
\begin{equation}
k_{x}^{2}+k_{y}^{2}+\bar{\alpha}^{2}=(k_{y}^{\prime }-if)(k_{y}^{\prime
}+if),  \label{22}
\end{equation}
where 
\begin{equation}
f^{2}\equiv k_{x}^{2}+\bar{\alpha}^{2}.  \label{23}
\end{equation}
We see that $I_{1}$ and $I_{2}$ have poles at $k_{y}^{\prime }=\pm if$ ,and
at $k_{y}^{\prime }=k_{y}$ , where $I_{1}=I_{2}^{*}(a_{2}\rightarrow a_{1})$%
. These integrals can be done using the contours shown in Fig. 14. The
results are 
\begin{equation}
I_{1}=\pi \left\{ \frac{k_{x}}{f}\frac{1}{(f-ik_{y})}e^{-a_{1}(f-ik_{y})}-%
\frac{k_{x}}{(k_{x}^{2}+k_{y}^{2}+\bar{\alpha}^{2})}\right\} ,  \label{24}
\end{equation}
and 
\begin{equation}
I_{2}=\pi \left\{ \frac{k_{x}}{f}\frac{1}{(f+ik_{y})}e^{-a_{2}(f+ik_{y})}-%
\frac{k_{x}}{(k_{x}^{2}+k_{y}^{2}+\bar{\alpha}^{2})}\right\} .  \label{25}
\end{equation}
Substituting these expressions for $I_{1\text{ }}$and $I_{2}$ into Eq. (19)
above, one obtains Eqs.(2). Eq.(3), the expression for $E_{y}(k_{x},k_{y})$
can be derived in a similar manner. Note that when $a_{1,2}\rightarrow 0$,
Eq. (19) gives the result for the transition radiation field as
expected.\pagebreak

\begin{description}
\item  $\mathbf{FigureCaptions}$

\item  \textbf{Figure 1}. A side view(1A), and a top view (1B), of \
backward diffraction radiation emitted by a charge passing through a slit
which is oriented with its edge parallel to the plane of incidence (\textbf{%
n, v} plane).

\item  \textbf{Figure 2}. A side view(2A), and a top view (2B), of backward
diffraction radiation emitted by a charge passing through a slit which is
oriented with its edge perpendicular to the plane of incidence (\textbf{n, v}
plane).

\item  \textbf{Figure 3. }Horizontal (3A) and vertical components (3B) of
the intensity of DR for $R=a/\gamma \bar{\lambda}=0.5$; the intensity (Z)
axis is scaled in proportion to the corresponding intensity component of TR.

\item  \textbf{Figure 4.} Horizontal (4A) and vertical components (4B) of
the intensity of DR for $R=2.0$; the intensity axis is scaled in proportion
to the corresponding intensity component of TR.

\item  \textbf{Figure 5}. Effect of the divergence $\delta ^{\prime }$ on
the horizontal component of the intensity. Shown are line scans in the
variable $X$ for two values of the variable $Y=\delta ^{\prime }=$ $0.2$ and 
$Y=\delta =0,$ for a fixed value of the divergence $\varepsilon ^{\prime
}=0.1$ and $R=2.0$.

\item  \textbf{Figure 6}. Effect of divergence $\varepsilon ^{\prime }$ on
the horizontal intensity. Shown are line scans in the variable $X$ for a
fixed value of the beam size $\delta =300\mu $ and $R=2$.

\item  \textbf{Figure 7}. Effect of beam size $\delta $ on the horizontal
intensity. Shown are line scans taken in the $X,Z$ plane, i.e. $Y=0$, for a
fixed value of divergence $\varepsilon ^{\prime }=0$ and $R=2.0$.

\item  \textbf{Figure 8}. Effect of the beam divergence $\varepsilon
^{\prime }$ on the vertical intensity. Shown are line scans in the variable $%
Y$ for two values of the variable $X=\varepsilon ^{\prime }=0.2$ and $%
X=\varepsilon ^{\prime }=0$, for a fixed value of the divergence $\delta
^{\prime }=0.1$, and $R=2$.

\item  \textbf{Figure 9}. Effect of divergence $\delta ^{\prime }$ on the
vertical intensity. Shown are line scans taken at $X=0$ for a fixed value of
the beam size, $\delta =$200$\mu $ and $R=2.0$.

\item  \textbf{Figure 10}. Effect of beam size $\delta $ on the vertical
intensity. Shown are line scans taken at $X=0$ for a fixed value of the
divergence, $\delta ^{\prime }=0$ and $R=2.0$.

\item  \textbf{Figure 11}. Configuration of a DR interferometer composed of
two slits oriented with the edges parallel to the plane of incidence.

\item  \textbf{Figure12}. Vertical component of intensity from a DR
interferometer for an electron beam with energy $E=500$ MeV; $R=2$, $\lambda
=3.2\mu m$, $\delta =200\mu $, $\delta ^{\prime }=0.05$ and separation
distance $L=6L_{v}=3$ meters.

\item  \textbf{Figure13. }Horizontal component of intensity from a DR
interferometer for an electron beam with same parameters as in Fig. 12.

\item  \textbf{Figure 14}. Contours used to evaluate the integrals $I_{1}$
and $I_{2}$ defined by Eqs. (20) and (21), respectively.mo

\newpage\centerline{\epsfxsize=\textwidth\epsfbox{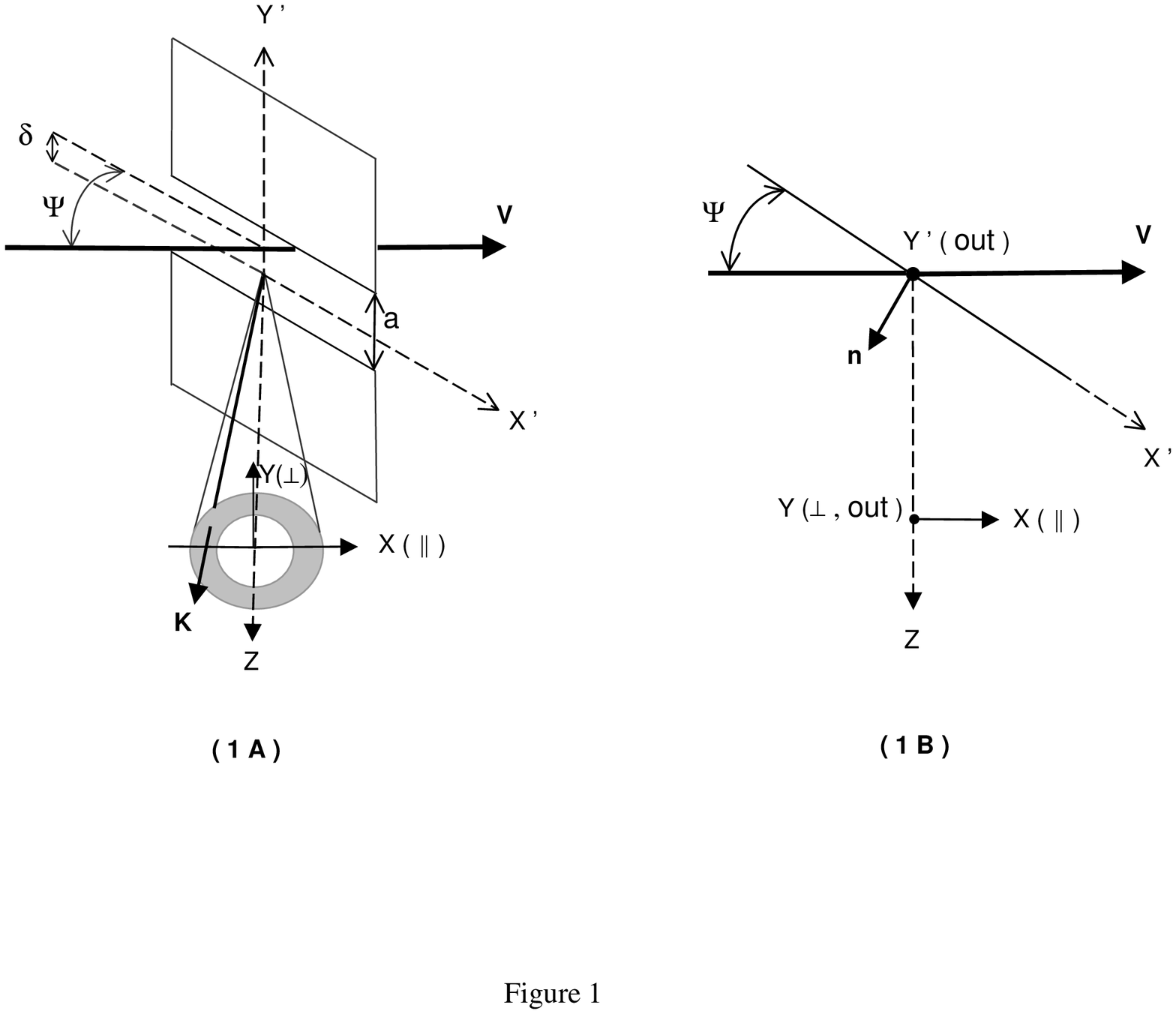}}
\newpage\centerline{\epsfxsize=\textwidth\epsfbox{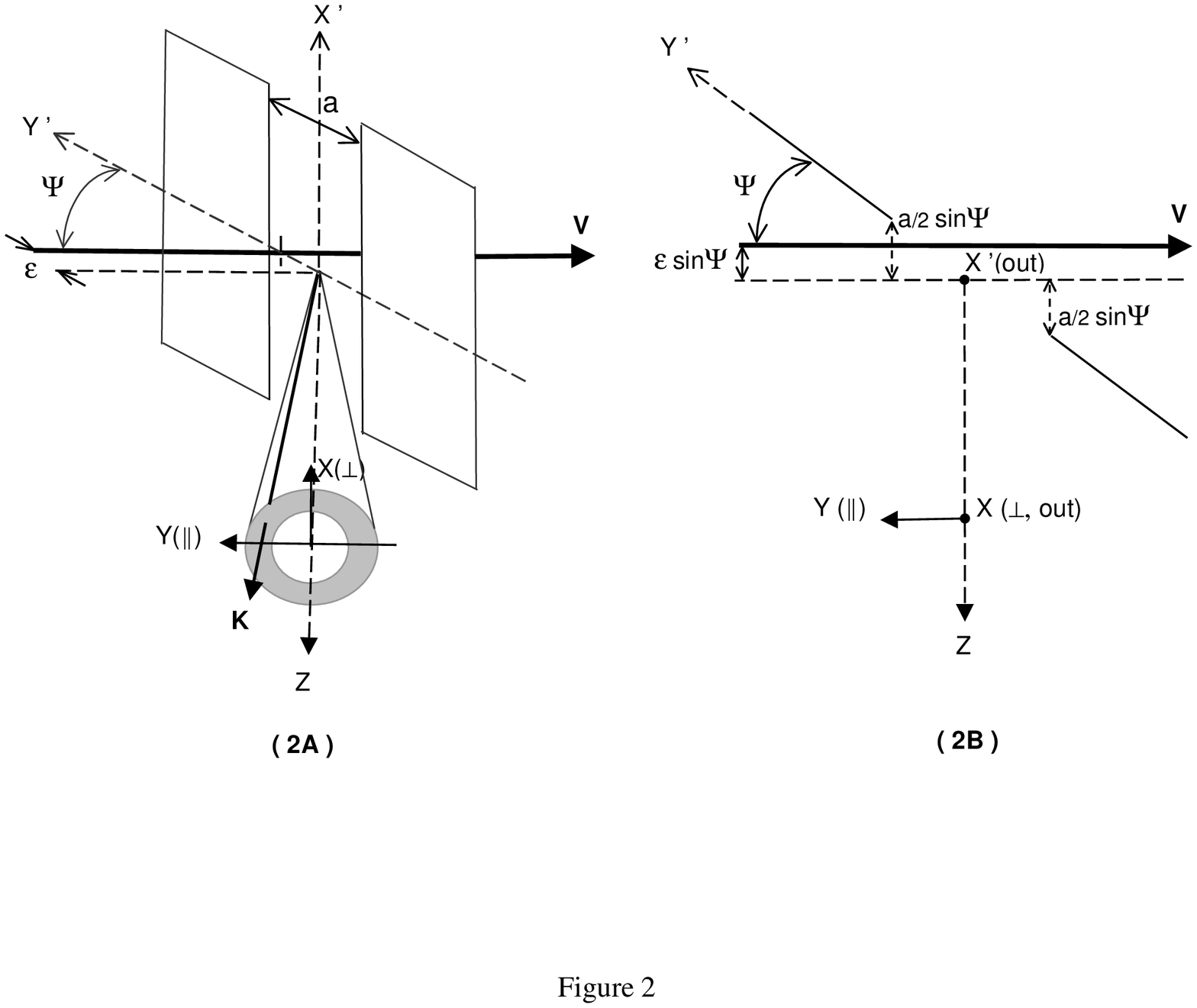}}
\newpage\centerline{\epsfxsize=\textwidth\epsfbox{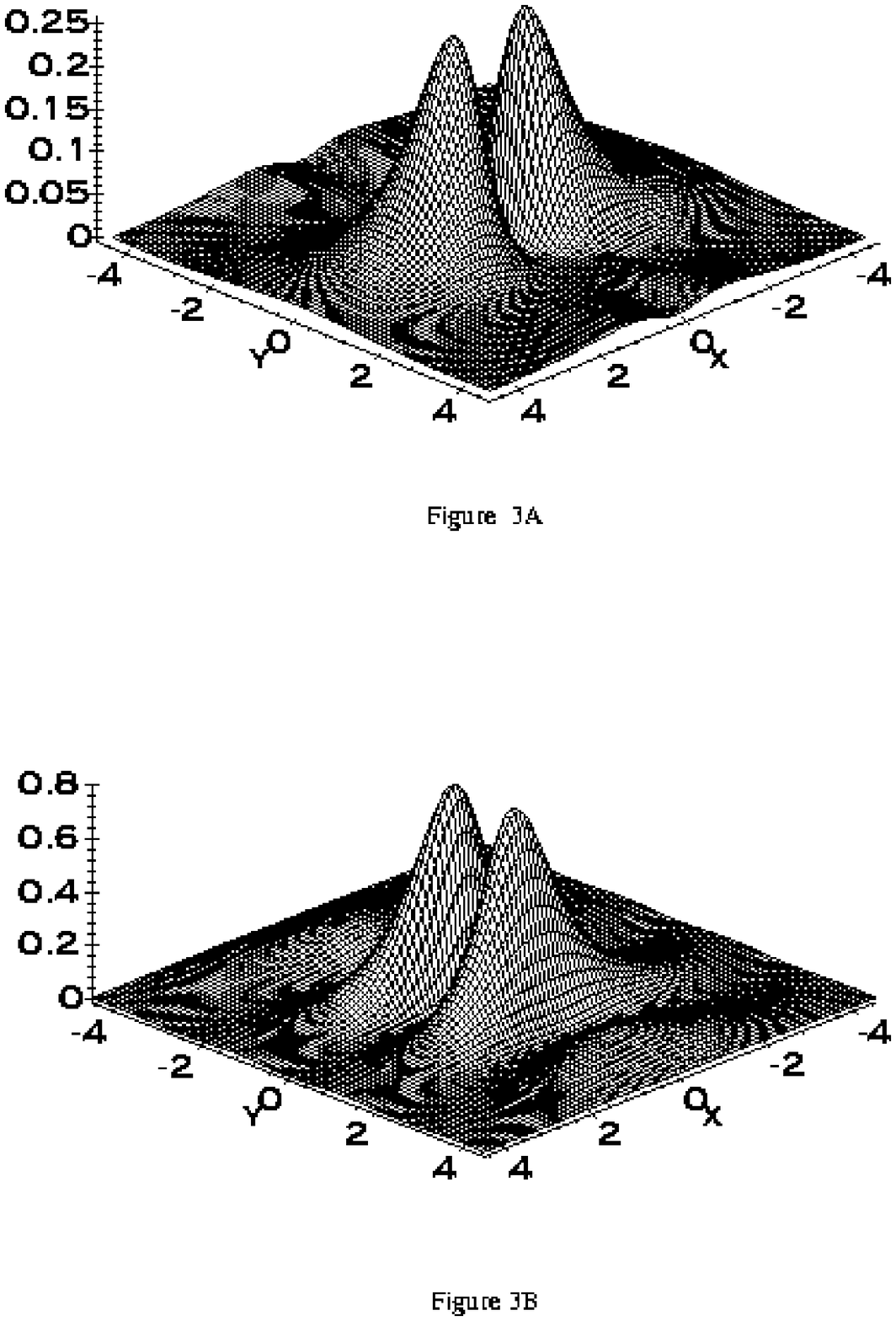}}
\newpage\centerline{\epsfxsize=\textwidth\epsfbox{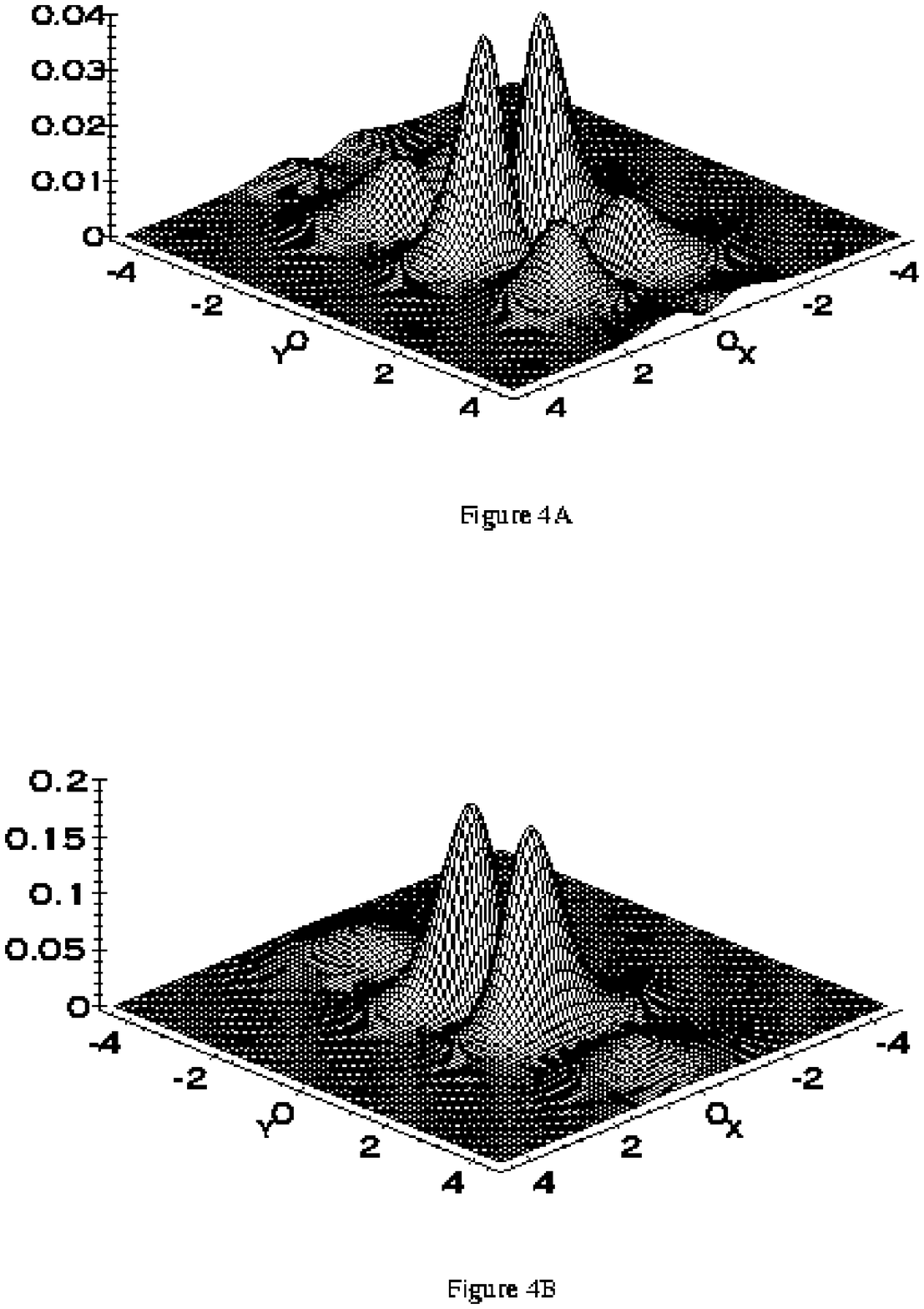}}
\newpage\centerline{\epsfxsize=\textwidth\epsfbox{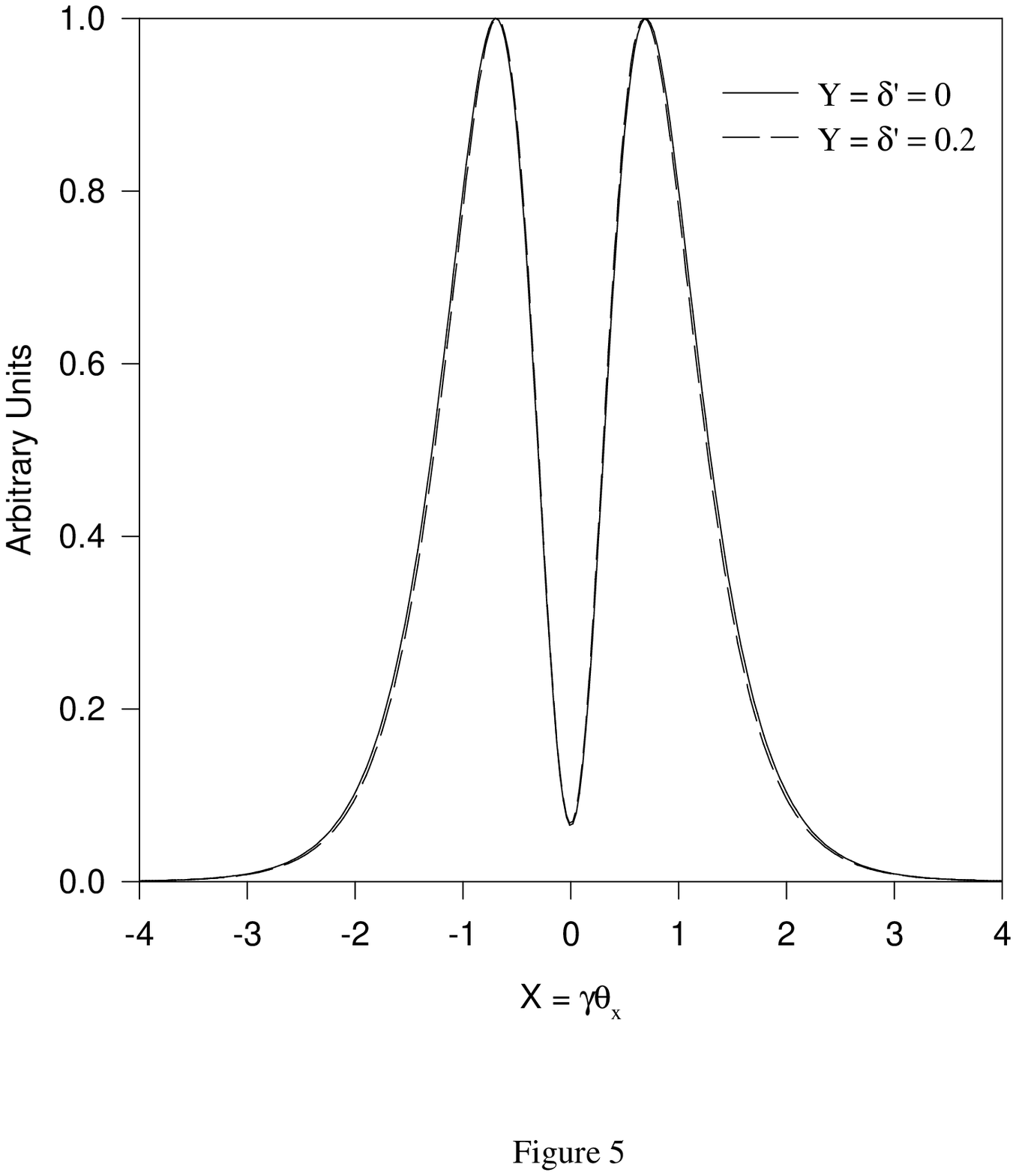}}
\newpage\centerline{\epsfxsize=\textwidth\epsfbox{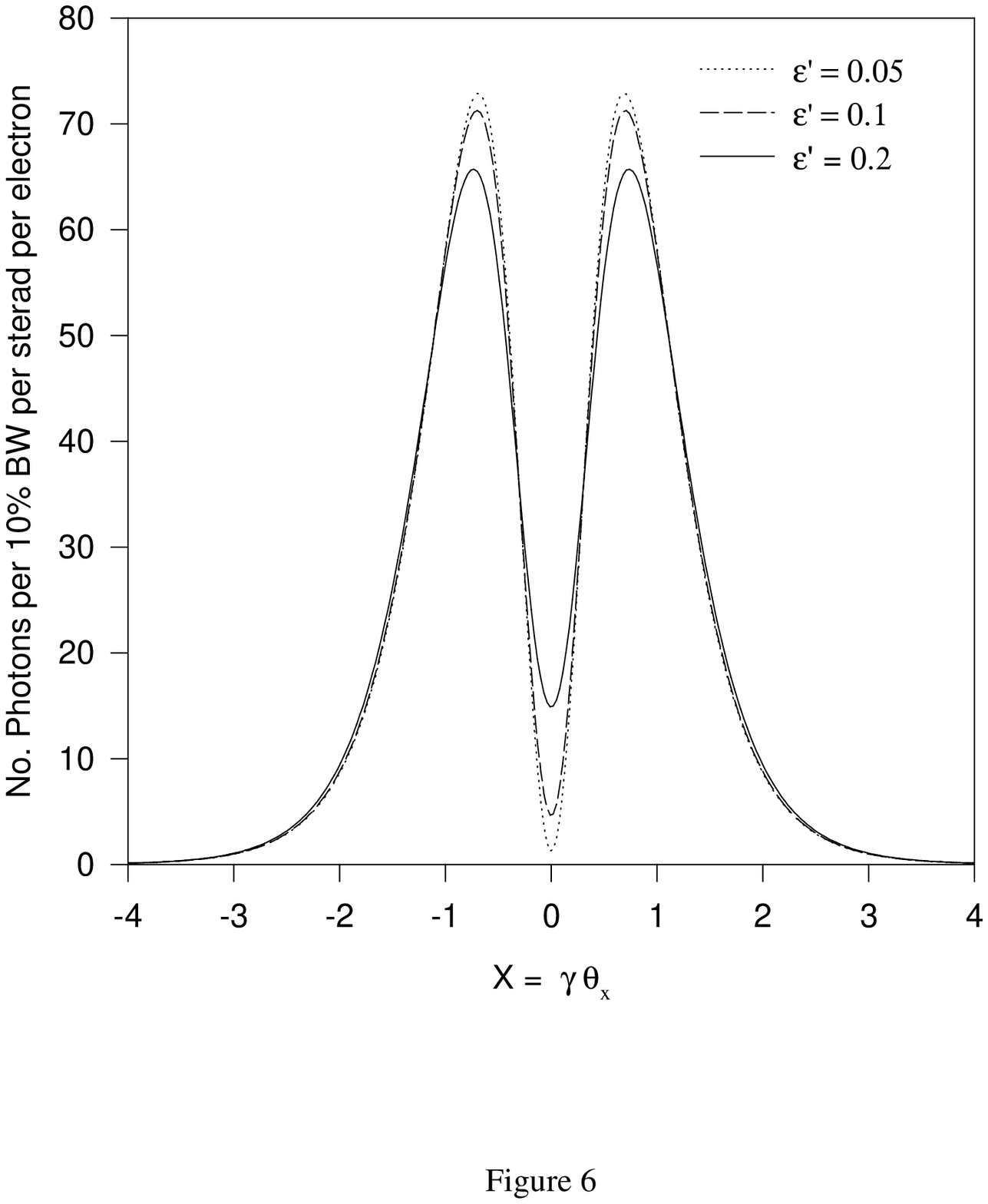}}
\newpage\centerline{\epsfxsize=\textwidth\epsfbox{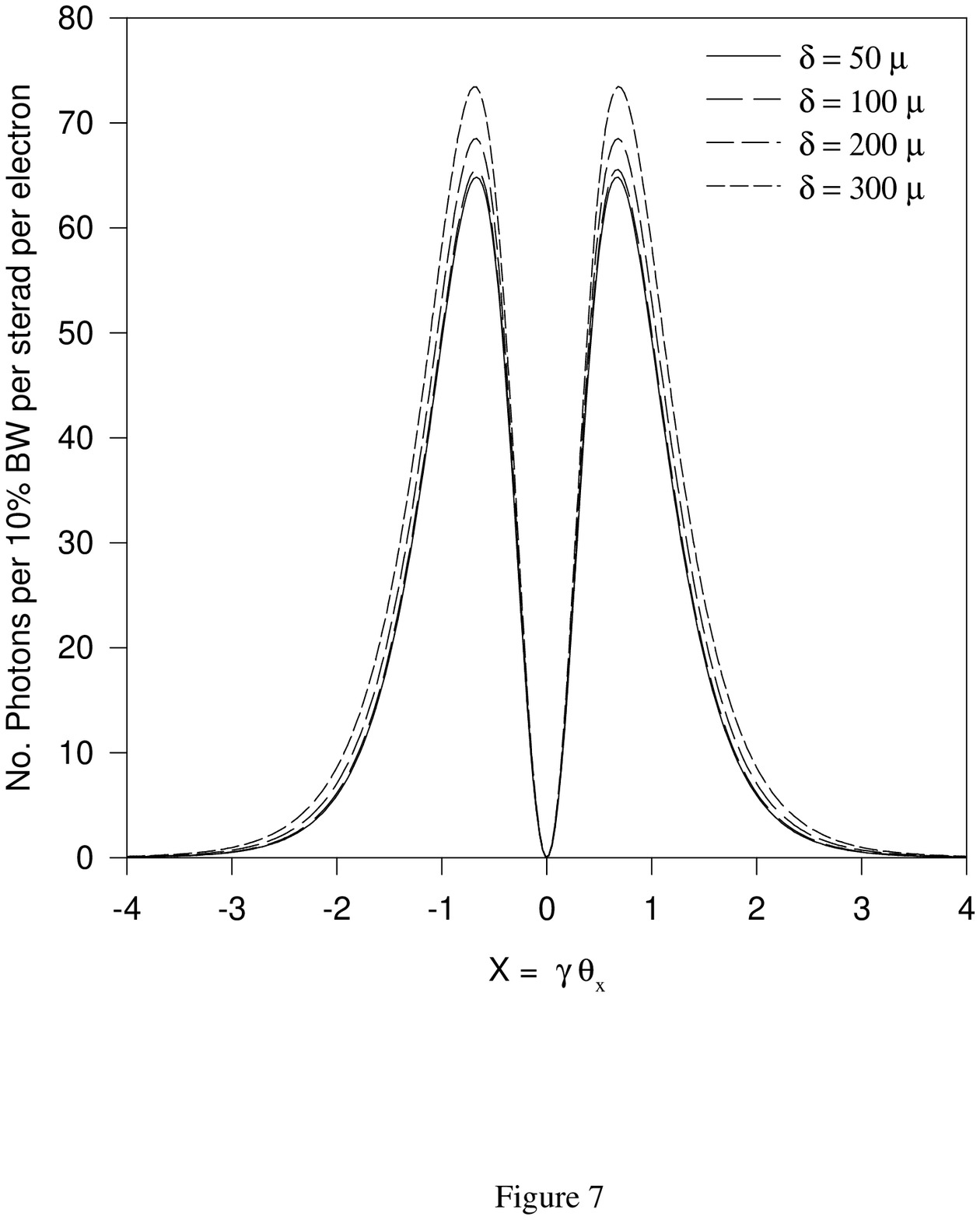}}
\newpage\centerline{\epsfxsize=\textwidth\epsfbox{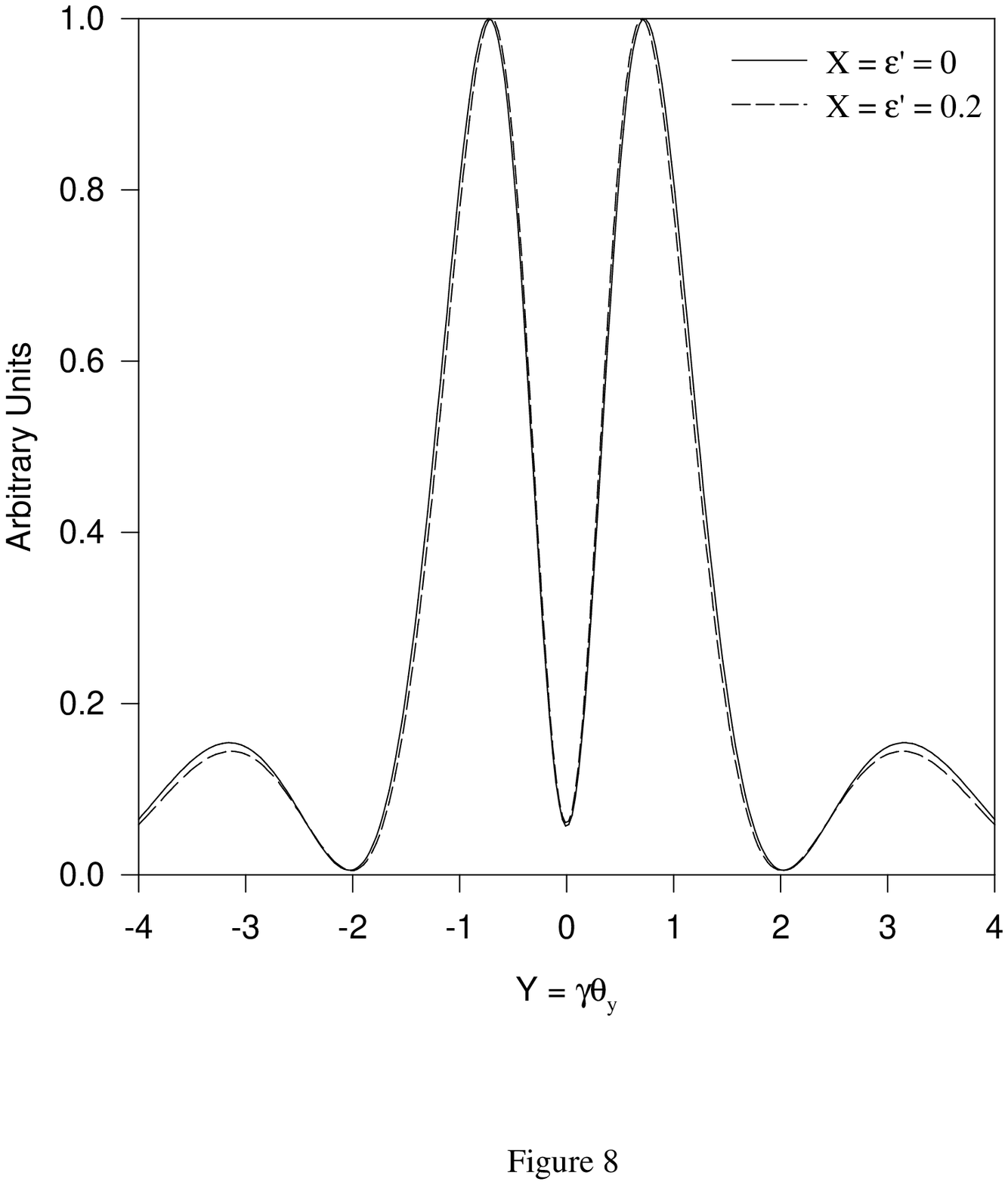}}
\newpage\centerline{\epsfxsize=\textwidth\epsfbox{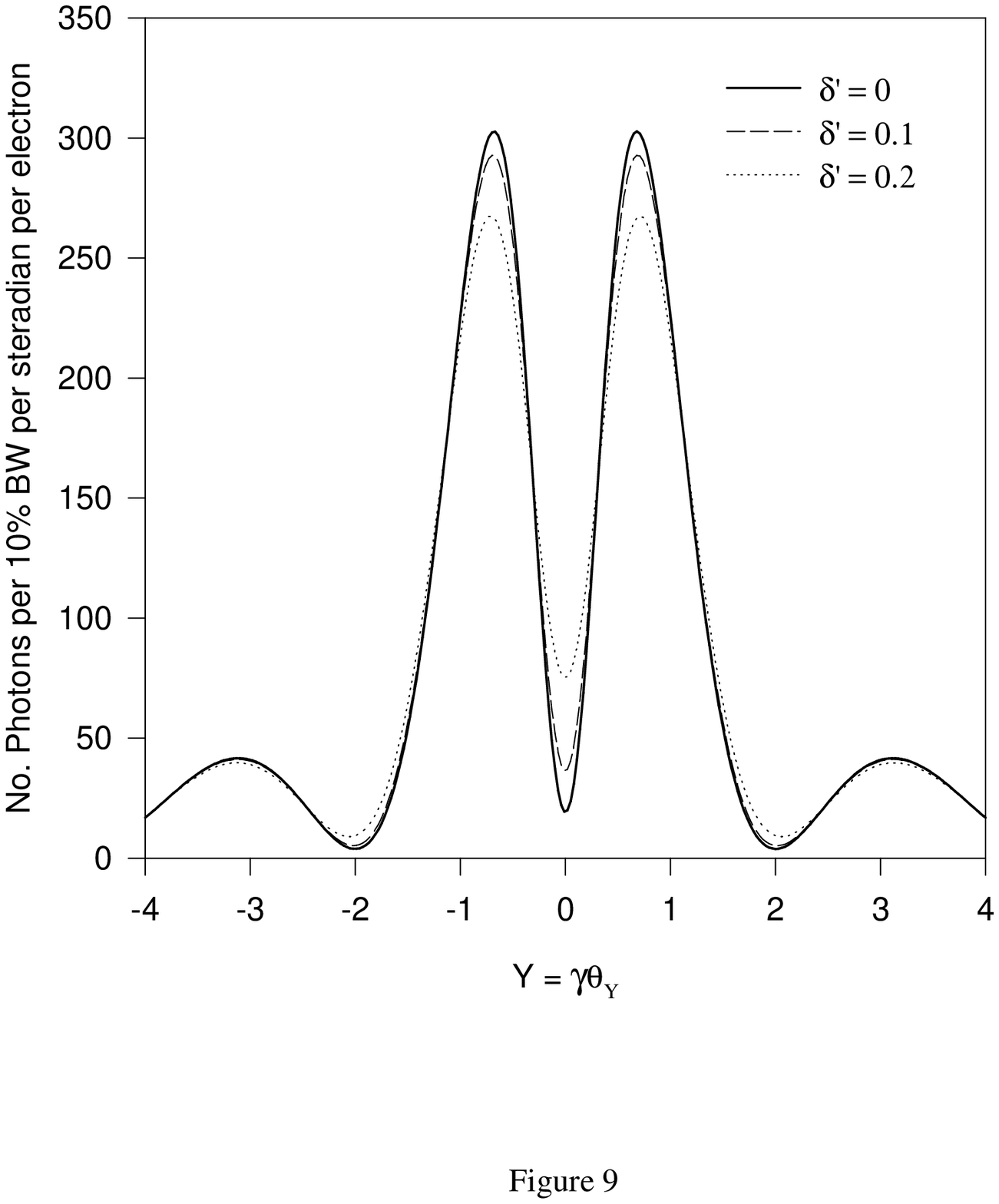}}
\newpage\centerline{\epsfxsize=\textwidth\epsfbox{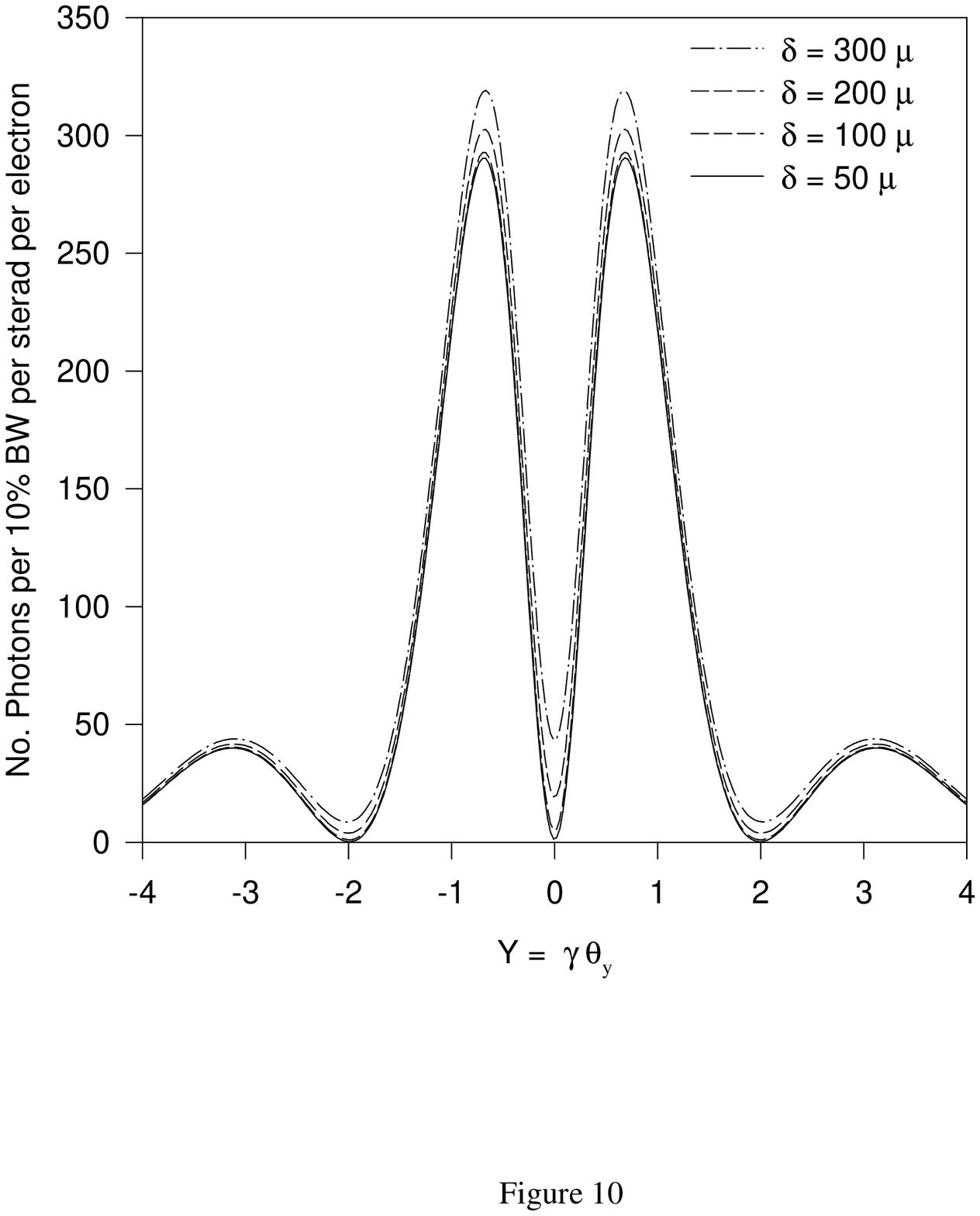}}
\newpage\centerline{\epsfxsize=\textwidth\epsfbox{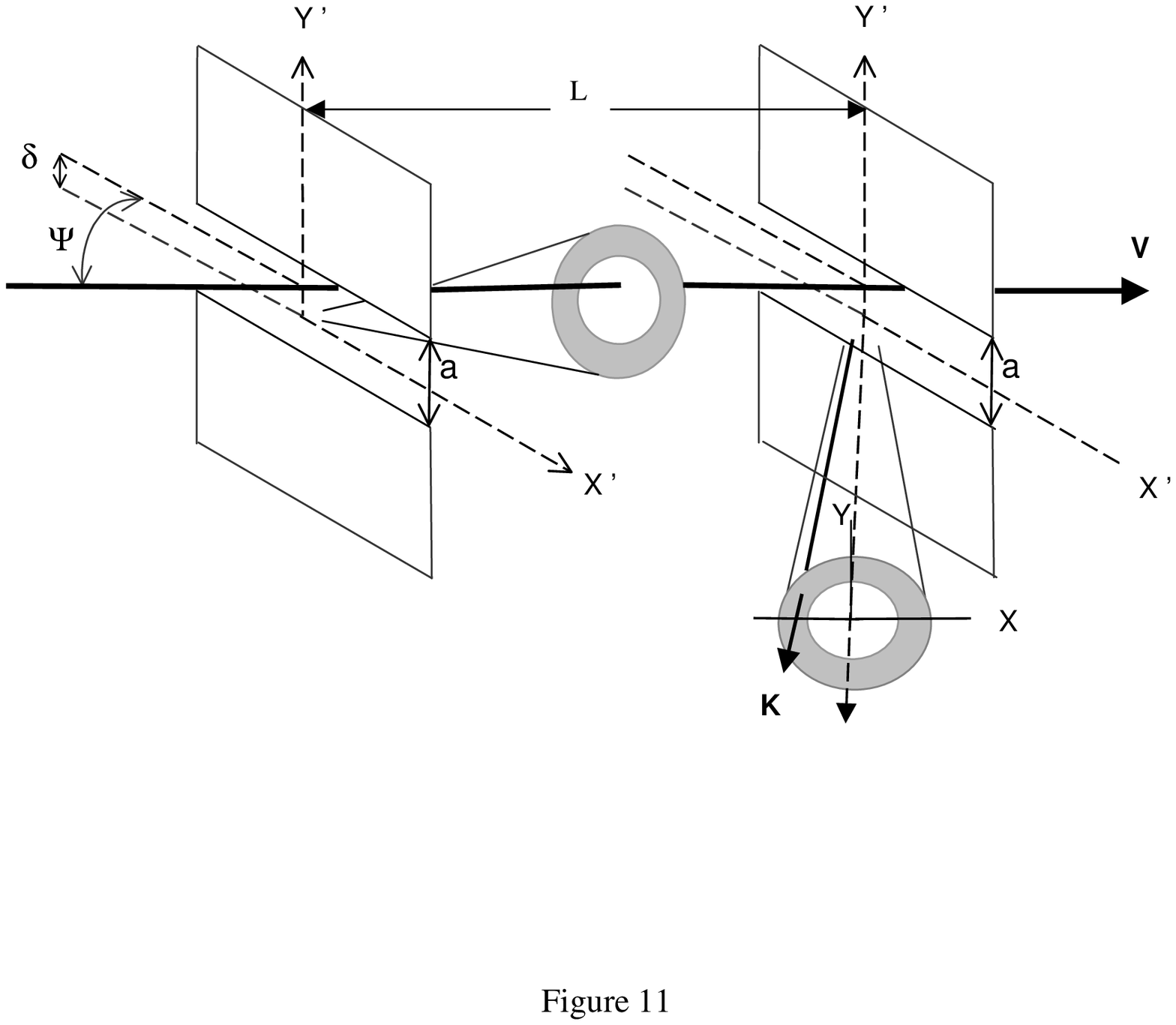}}
\newpage\centerline{\epsfxsize=\textwidth\epsfbox{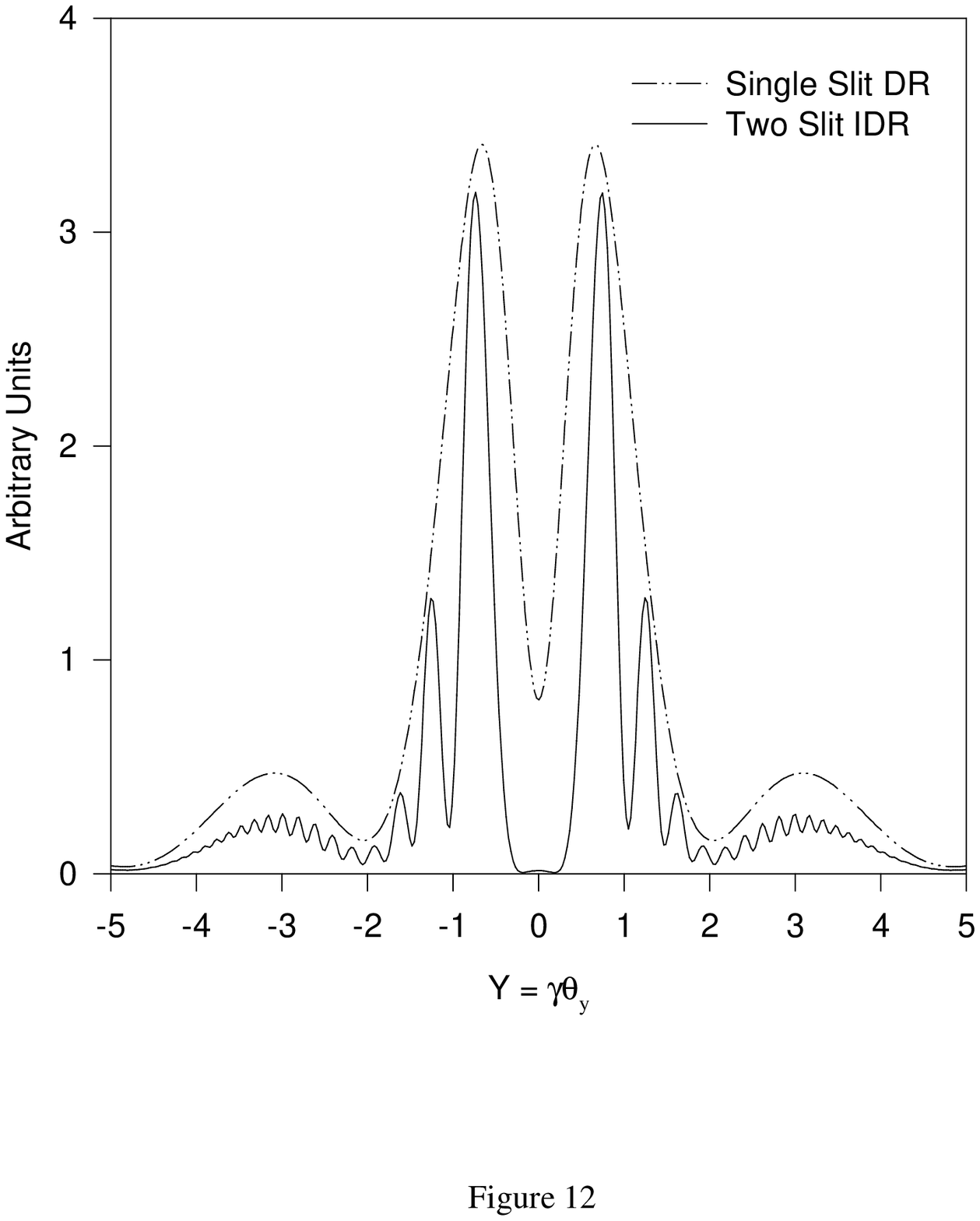}}
\newpage\centerline{\epsfxsize=\textwidth\epsfbox{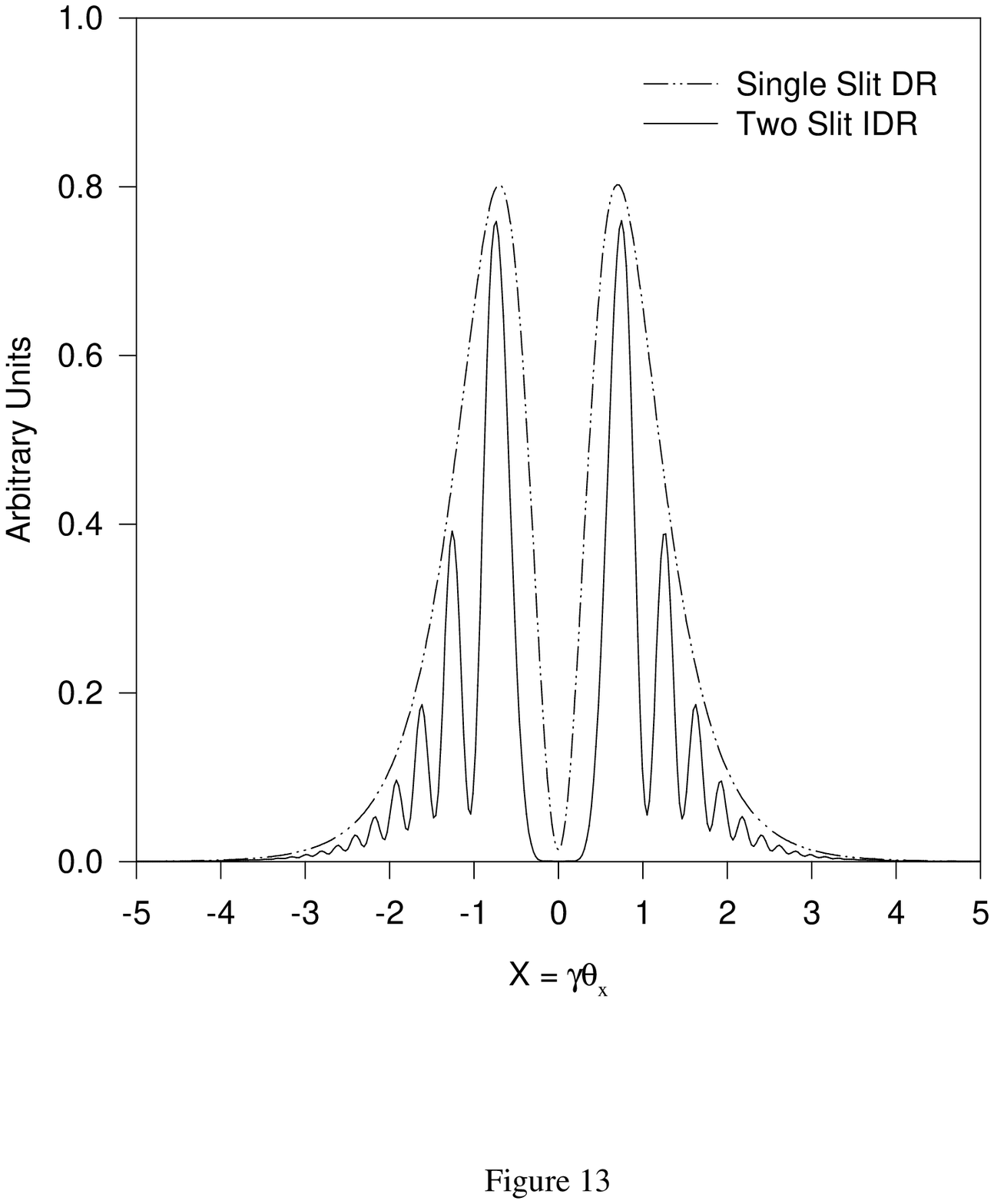}}
\newpage\centerline{\epsfxsize=\textwidth\epsfbox{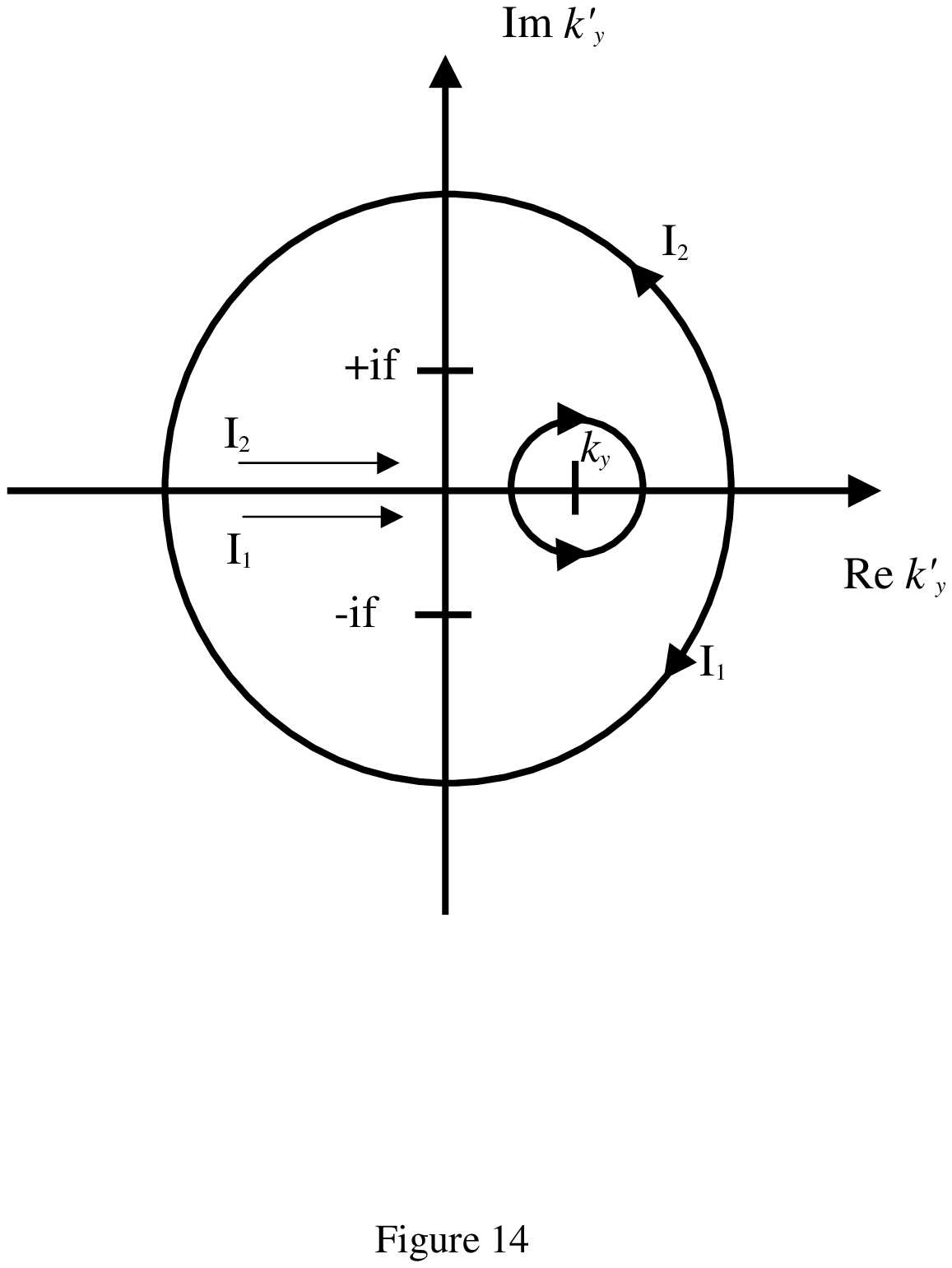}}

\end{description}
\end{document}